\documentclass[fleqn,usenatbib]{mnras}

\usepackage{newtxtext,newtxmath}

\usepackage[T1]{fontenc}
\usepackage{ae,aecompl}

\usepackage{graphicx}	
\usepackage{amsmath}	
\usepackage{multicol} 
\usepackage{multirow}
\usepackage{comment}
\usepackage{array}
\usepackage{soul}
\hypersetup{colorlinks=true,citecolor=blue,linkcolor=purple,filecolor=black,runcolor=black,breaklinks=true}
\usepackage{etoolbox}
\usepackage{ae,aecompl}
\usepackage[usenames]{color}
\usepackage{hyperref}
\usepackage{threeparttable}
\usepackage{mathrsfs}
\usepackage{scalerel}
\usepackage{longtable,booktabs,threeparttablex}

\definecolor{purple}{RGB}{160,32,240}

\definecolor{red}{RGB}{225,50,50}

\newcommand{\HST}{\emph{HST}}
\newcommand{\JWST}{\emph{JWST}}
\newcommand{\Spitzer}{\emph{Spitzer}}
\newcommand{\Muv}{\ensuremath{\mathrm{M}_{\mathrm{UV}}^{ }}}

\newcommand{\logten}{\ensuremath{\log_{10}}}

\newcommand{\Lya}{\ensuremath{\mathrm{Ly}\alpha}}
\newcommand{\zLya}{\ensuremath{\mathrm{z}_{_{\mathrm{Ly\alpha}}}}}

\newcommand{\zsys}{\ensuremath{\mathrm{z}_{\mathrm{sys}}}}

\newcommand{\fesc}{\ensuremath{f_{\mathrm{esc}}}}

\newcommand{\muEW}{\ensuremath{\mu_{\mathrm{EW}}}}
\newcommand{\sigmaEW}{\ensuremath{\sigma_{\mathrm{EW}}}}

\newcommand{\Msol}{\ensuremath{\mathrm{M}_{\odot}}}
\newcommand{\Mstar}{\ensuremath{\mathrm{M}_{\ast}}}
\newcommand{\logMstar}{\ensuremath{\log_{10}\left(\mathrm{M}_{\ast}/\mathrm{M}_{\odot}\right)}}

\newcolumntype{P}[1]{>{\centering\arraybackslash}p{#1}}
\newcommand\Tstrut{\rule{0pt}{2.6ex}}         
\newcommand\Bstrut{\rule[-1.2ex]{0pt}{0pt}}   

\newcommand{\rbracket}{]}

\title[The [OIII\rbracket{}$+$H$\beta$ Equivalent Width Distribution at z$\simeq$7]{The [OIII]$+$H$\beta$ Equivalent Width Distribution at z$\simeq$7: Implications for the Contribution of Galaxies to Reionization}

\author[R. Endsley et al.]{
Ryan Endsley$^{1}$\thanks{E-mail: rendsley@email.arizona.edu},
Daniel P. Stark$^{1}$, 
Jacopo Chevallard$^{2}$, \& 
St\'ephane Charlot$^{2}$
\\
$^{1}$Steward Observatory, University of Arizona, 933 N Cherry Ave, Tucson, AZ 85721 USA\\
$^{2}$Sorbonne Universit\'es, UPMC-CNRS, UMR7095, Institut dAstrophysique de Paris, F-75014, Paris, France
}

\date{Accepted XXX. Received YYY; in original form ZZZ}

\pubyear{2020}

\begin{document}
\label{firstpage}
\pagerange{\pageref{firstpage}--\pageref{lastpage}}
\maketitle


\begin{abstract}
We quantify the distribution of [OIII]+H$\beta$ line strengths at z$\simeq$7 using a sample of 20 bright (\Muv{}~$\lesssim$~-21) galaxies. 
We select these systems over wide-area fields (2.3 deg$^2$ total) using a new colour-selection which precisely selects galaxies at z$\simeq$6.63--6.83, a redshift range where blue \Spitzer{}/IRAC [3.6]$-$[4.5] colours unambiguously indicate strong [OIII]$+$H$\beta$ emission. 
These 20 galaxies suggest a log-normal [OIII]$+$H$\beta$ EW distribution with median EW = 759$^{\scaleto{+112}{4.5pt}}_{\scaleto{-113}{4.5pt}}$~\AA{} and standard deviation = 0.26$^{\scaleto{+0.06}{4.5pt}}_{\scaleto{-0.05}{4.5pt}}$ dex. 
We find no evidence for strong variation in this EW distribution with UV luminosity.
The typical [OIII]+H$\beta$ EW at z$\simeq$7 implied by our sample is considerably larger than that in massive star forming galaxies at z$\simeq$2, consistent with a shift toward larger average sSFR (4.4 Gyr$^{-1}$) and lower metallicities (0.16 Z$_\odot$). 
We also find evidence for the emergence of a population with yet more extreme nebular emission ([OIII]+H$\beta$ EW$>$1200~\AA{}) that is rarely seen at lower redshifts. 
These objects have extremely large sSFR ($>$30 Gyr$^{-1}$), as would be expected for systems undergoing a burst or upturn in star formation. 
While this may be a short-lived phase, our results suggest that 20\% of the z$\simeq$7 population has such extreme nebular emission, implying that galaxies likely undergo intense star formation episodes regularly at z$>$6. 
We argue that this population may be among the most effective ionizing agents in the reionization era, both in terms of photon production efficiency and escape fraction.
We furthermore suggest that galaxies passing through this large sSFR phase are likely to be very efficient in forming bound star clusters.  

\end{abstract}




\section{Introduction}

Over the past decade, deep imaging surveys have begun to unveil galaxies present at z$>$6 (e.g. \citealt{McLure2013,Bowler2014,Atek2015b,Bouwens2015_LF,Finkelstein2015_LF,Livermore2017,Oesch2018_z10LF,Ono2018}), providing our first census of star formation in the first billion years. 
These galaxies are found to have lower stellar masses, higher specific star formation rates (sSFRs), and bluer rest-UV continuum slopes than those typically found at lower redshifts (see \citealt{Stark2016_ARAA} for a review). 
The ionizing output of this population is in principle sufficient for galaxies to drive reionization, provided that a large enough fraction of Lyman Continuum (LyC) radiation (10-20\%) is able to escape into the intergalactic medium (IGM) \citep{Bouwens2015_reionization,Ishigaki2015,Robertson2015,Finkelstein2019}. 
Unfortunately, little is  known about the ionizing efficiency of early galaxies, making it difficult to verify whether this condition is met.

With the launch of the {\it James Webb Space Telescope (JWST)}, attention will soon start to focus on the spectra of early galaxies, providing a path toward characterizing their effectiveness as ionizing agents. 
Our first glimpse of the emission line properties has been made possible by deep {\it Spitzer}/IRAC imaging. At 6$<$z$<$9, the rest-frame optical lines are situated in the IRAC 3.6$\mu$m and 4.5$\mu$m broad-band filters.  
Galaxies at z$\simeq$8 have been shown to have atypically red [3.6]$-$[4.5] IRAC colours, as expected if the [4.5] filter has a contribution from [OIII] and H$\beta$ emission. 
The average IRAC colours point to intense [OIII]$+$H$\beta$ emission with rest-frame equivalent width (EW) of 650--670 \AA{} in typical systems at z$\simeq$8 \citep{Labbe2013,deBarros2019}. 
Similar findings appear for galaxies at slightly lower redshifts (6.6$<$z$<$6.9) where [OIII] and H$\beta$ fall in the [3.6] bandpass \citep{Smit2014,Smit2015}.  
Individual galaxies have been identified with yet more extreme IRAC colours pointing to [OIII]$+$H$\beta$ emission with EW=1000--2000~\AA\ \citep{Smit2014,Smit2015,RobertsBorsani2016}. 
Such extreme EW [OIII]+H$\beta$ emission ($>$1000~\AA) arises when the light from very young (1--10 Myr) stellar populations dominates the emergent spectrum in the UV and optical (e.g. \citealt{Tang2019}), as expected for galaxies that have recently formed a large number of dense massive star clusters in a recent upturn of star formation \citep{Tang2019,Vanzella2020}.
Spectroscopic follow-up of this z$>$7 population has revealed intense nebular emission from highly ionized species of carbon in the rest-UV \citep{Stark2015_CIV,Stark2015_CIII,Stark2017,Laporte2017,Mainali2017,Schmidt2017,Hutchison2019}, suggesting extremely efficient ionizing production may be relatively common in the subset of early galaxies with [OIII]+H$\beta$ EW $>$ 1000--2000~\AA. 

The ubiquity of such extremely large ($>$1000~\AA) EW [OIII]+H$\beta$ emitters remains a matter of debate at z$>$7. 
While several studies have argued that these systems are relatively common \citep{Smit2014,Smit2015,deBarros2019}, others have cast doubt on this conclusion following the discovery of Balmer Breaks (and hence old stellar populations)
in several galaxies at redshifts as high as z$\simeq$9 (e.g. \citealt{Hashimoto2018_z9}; \citealt{Strait2020}). 
It has additionally been noted that the IRAC [4.5] flux excesses which are typically linked to [OIII]+H$\beta$ emission at z$>$7 (e.g. \citealt{RobertsBorsani2016}) could also arise from strong Balmer breaks \citep{RobertsBorsani2020}, creating yet more uncertainty in the fraction of the population caught in such an extreme line emitting phase.  

In this paper, we characterize the [OIII]+H$\beta$ EW distribution in z$\simeq$7 galaxies, with the primary goal of assessing the fraction of the population caught in the most extreme line emitting phase ([OIII]+H$\beta$ EW $>$1000~\AA) where galaxies appear to be effective ionizing agents.  
We focus our study in the redshift range 6.6$<$z$<$6.9 to avoid the degeneracy between Balmer Breaks and nebular emission that plagues z$>$7 studies \citep{RobertsBorsani2020}.
To reliably establish the [OIII]+H$\beta$ EW distribution, we need to address two challenges that have previously limited past works in this redshift range. 
First, the availability of only broad-band photometry in deep \HST{} fields makes it very difficult to photometrically identify only those sources that fall within the required narrow redshift window. 
With a broader redshift selection, the IRAC colours cannot be unambiguously linked to nebular emission. 
Second, there is generally a mismatch between the sensitivities of \Spitzer{} and \HST{}, making it difficult to infer [OIII]+H$\beta$ EWs for the faint z$\sim$7 galaxies detected with \HST{}. 

Here we overcome both challenges by investigating galaxies in ground-based imaging fields with deep IRAC photometry. 
The wide areas of these fields enables us to build sizeable sample of 
bright (\Muv{} $\lesssim$ $-$21) galaxies that are well-matched in magnitude to the 
existing IRAC depths. 
By exploiting imaging from four narrow-band and broad-band filters over these fields, we are able to develop a new \Lya{}-break dropout selection that precisely picks out galaxies in the narrow redshift range (6.6$<$z$<$6.9) where the [OIII]+H$\beta$ EW can be reliably inferred from IRAC photometry. 
Using this sample, we derive a functional form for the [OIII]+H$\beta$ EW distribution at z$\simeq$7.
We also supplement this new bright galaxy sample with an \HST{}-based selection, enabling us to probe fainter (\Muv{} $\lesssim$ $-$20) and test whether the [OIII]$+$H$\beta$ EW distribution evolves strongly with UV luminosity at z$\simeq$7.  With these distributions in hand, we quantify the fraction of the early galaxy population caught in the very extreme emission line phase ([OIII]+H$\beta$ EW $>$1000--2000~\AA) that appears linked to efficient ionizing photon production and escape.

The paper is organized as follows. 
We detail our selection criteria in \S\ref{sec:sample_selection} and report the inferred properties (e.g. stellar mass, sSFR, and [OIII]$+$H$\beta$ EW) of each galaxy in \S\ref{sec:galaxy_properties}.
In \S\ref{sec:spec_confirmation}, we report a new \Lya{} detection from an extremely luminous (\Muv{} = $-$22.5) galaxy at z=6.850, and use the spectroscopic redshift to better constrain the inferred [OIII]$+$H$\beta$ EW.
We infer the [OIII]$+$H$\beta$ EW distribution at z$\simeq$7 in \S\ref{sec:analysis}, deriving both the median EW and the standard deviation.
In \S\ref{sec:discussion}, we then discuss implications for the efficiency of ionizing photon production and escape in the reionization era. 
Our main conclusions are summarized in \S\ref{sec:summary}.

All magnitudes are quoted in the AB system \citep{OkeGunn1983} and we adopt a flat $\Lambda$CDM cosmology with h=0.7, $\Omega_\mathrm{M}$ = 0.3, and $\Omega_\mathrm{\Lambda}$ = 0.7, consistent with \textit{Planck} results \citep{Planck2018}.


\section{Observations} \label{sec:Observations}

We describe our selection criteria and the resulting photometric properties of our sample in \S\ref{sec:sample_selection}. We then infer the galaxy properties (e.g. stellar mass and [OIII]+H$\beta$ EW) of each source in \S\ref{sec:galaxy_properties} using a photoionization model SED fitting code, BEAGLE \citep{Chevallard2016}. In \S\ref{sec:spec_confirmation}, we report results from our first spectroscopic observations of this sample aimed at targeting \Lya{} in one of the most luminous and strongest [OIII]+H$\beta$ emitters in our sample.

\subsection{Sample Selection} \label{sec:sample_selection}

We seek to infer the ionizing properties of luminous ($-$22 $\lesssim$ \Muv $\lesssim$ $-$20) reionization-era galaxies. While direct spectroscopic measurements of strong rest-optical lines at z$>$6.5 are currently not possible, we can infer their strength with \Spitzer{}/IRAC colours. Specifically, at z$\simeq$6.63--6.83, H$\beta$ and [OIII]$\lambda$4959,5007 lie within the 3.6$\mu$m IRAC filter while the 4.5$\mu$m filter is free of strong lines.\footnote{Here, we are quoting redshifts where H$\beta$ and both components of the [OIII]$\lambda$4959,5007 doublet lie at wavelengths of $>$50\% maximum transmission through the 3.6$\mu$m filter and H$\alpha$ is at $<$50\% maximum transmission through the 4.5$\mu$m filter} Therefore, strong rest-optical line emitters over this redshift interval will show a blue [3.6]$-$[4.5] colour \citep[e.g.][]{Wilkins2013,Smit2014,Smit2015}. 

We choose to select z$\simeq$6.63--6.83 sources using both wide-area ground-based imaging as well as deep \HST{} imaging. By exploiting both, we are able to simultaneously capture the most luminous (\Muv{} $\lesssim$ $-$21) sources while also probing further down the luminosity function (to \Muv{} $\sim$ $-$20). Because photometric filter availability varies, our ground and \HST{}-based samples necessitate different selection criteria which we explain in turn below.

\begin{figure}
\includegraphics{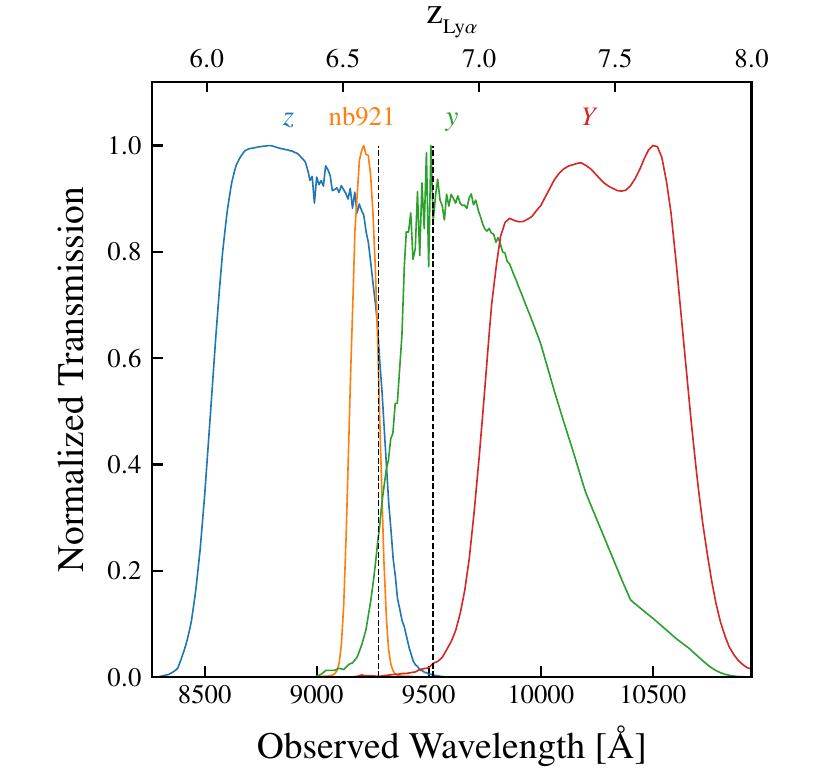}
\caption{Normalized transmission functions of the Subaru Hyper Suprime-Cam \textit{z}, \textit{y}, and nb921 filters as well as the VISTA/VIRCam \textit{Y} filter plotted as a function of wavelength. We also show in the top axis the redshift corresponding to \Lya{} (1215.67 \AA{}) at each wavelength. The vertical dashed black lines encompass the redshift range \zLya{} = 6.63--6.83 where young galaxies with strong rest-optical line emission would show blue [3.6]$-$[4.5] \Spitzer{}/IRAC colours. Galaxies at z=6.63--6.83 can be selected by their strong \textit{z} and nb921 drops with relatively flat \textit{y}-\textit{Y} colours.}
\label{fig:filters}
\end{figure}

\subsubsection{Ground-Based Selection over COSMOS and XMM1} \label{sec:COSMOSXMM1_Selection}

We perform ground-based selection of z$\simeq$6.63--6.83 sources over the COSMOS and XMM1 fields as both uniquely possess Subaru/Hyper Suprime-Cam (HSC) \textit{z}, nb921, and \textit{y} photometry \citep{Aihara2019} as well as VISTA/VIRCam \textit{Y} photometry \citep{McCracken2012,Jarvis2013}. As shown in Fig. \ref{fig:filters}, these four filters well sample the \Lya{}-break from z$\approx$6.5--7.0 and therefore enable us to develop selection criteria that, for the first time, precisely target the narrow redshift range of z$\simeq$6.63--6.83. At z$\geq$6.63, only a small fraction of flux redwards of the \Lya{}-break will fall within the \textit{z} and nb921 filters, while $>$75\% of the \textit{y}-band integrated transmission will contain flux redwards of the break out to z=6.83. Therefore, z=6.63--6.83 sources will appear as strong \textit{z} and nb921 dropouts yet have relatively flat \textit{y}-\textit{Y} colours. 

To determine specific colour cuts that optimally isolate z$\simeq$6.63--6.83 galaxies, we simulate the HSC and VIRCam colours of z=6--8 galaxies with a rest-UV slope (F$_{\nu} \propto \lambda^{\beta+2}$) of $\beta$ = $-$2 and adopt the IGM transmission function from \citet{Inoue2014}. We find that the following colour cuts best select z$\simeq$6.63--6.83 galaxies with this filter set:
\begin{enumerate}
    \item \textit{z}-\textit{y}$>$1.5
    \item \textit{z}-\textit{Y}$>$1.5
    \item nb921-\textit{Y}$>$1.0
    \item \textit{y}-\textit{Y}$<$0.4
\end{enumerate}
Fluxes in the \textit{z} and nb921 dropout bands are set to their 1$\sigma$ value in cases of non-detections, consistent with past literature \citep[e.g.][]{Bouwens2015_LF,Stefanon2019}. Here, we have assumed a \Lya{} EW = 0 \AA{}, motivated by the rapid drop in \Lya{} emitters observed at z$>$6.5 \citep[e.g.][]{Schenker2014}. Recent results have shown that even moderate ($>$25 \AA{} EW) \Lya{} emission is only seen in $\lesssim$10\% of bright (\Muv{} $<$ $-$20.25) z$\sim$7 galaxies \citep{Pentericci2018}. We discuss how \Lya{} emission would impact the redshift range of our selection below.

The COSMOS and XMM1 fields both further possess deep optical through mid-infrared imaging over a very large area of $\approx$1.5 and 0.8 deg$^2$, respectively, enabling additional selection criteria for a more robust sample. Specifically, both fields have been imaged from 0.4--1$\mu$m with HSC in the \textit{g},\textit{r},\textit{i},\textit{z},\textit{y} broad-bands as well as the nb816 and nb921 narrow-bands with the HSC Subaru Strategic Program (HSCSSP; \citealt{Aihara2018}). The typical 5$\sigma$ depth (1.2\arcsec{} diameter aperture) in COSMOS is m=27.7, 27.4, 27.3, 26.1, 26.9, 26.2, and 26.1 in \textit{g}, \textit{r}, \textit{i}, nb816, \textit{z}, nb921, and \textit{y}, respectively\footnote{Here we are using the most recent HSCSSP release, PDR2 \citep{Aihara2019}.}. In XMM1, the HSC depths are typically 0.5--1 mag shallower reaching m=27.3, 26.7, 26.8, 26.2, 26.1, 26.0, and 25.3, respectively. 

Furthermore, the UltraVISTA survey \citep{McCracken2012} provides near-infrared \textit{Y}, \textit{J}, \textit{H}, and \textit{K$_s$} broad-band imaging from VISTA/VIRCam over COSMOS. The UltraVISTA coverage is split roughly equally into deep and ultra-deep stripes. The deep stripes have typical 5$\sigma$ depths of m=24.7, 24.5, 24.2, and 24.5 in \textit{Y}, \textit{J}, \textit{H}, and \textit{K$_s$}, respectively. The depths in the ultra-deep stripes are typically $>$1 mag deeper reaching m=25.9, 25.8, 25.6, and 25.2, respectively. The ultra-deep stripes were also imaged with the VIRCam narrow-band filter NB118 to a 5$\sigma$ depth of m=24.5.

The VIDEO \citep{Jarvis2013} and UKIDSS \citep{Lawrence2007} surveys cover XMM1 with VISTA/VIRCam \textit{Y}, \textit{H}, \textit{K}$_s$ and UKIRT/WFCam \textit{J}, \textit{H}, \textit{K} fiters, respectively. For the purposes of this work, we ignore the VIRCam \textit{H} and \textit{K}$_s$ photometry over XMM1 as the WFCam \textit{H} and \textit{K} depths are typically $\approx$1 and 1.5 mags deeper, respectively. Specifically, the typical 5$\sigma$ depths are 25.2 in VIRCam \textit{Y} and 25.7, 25.0, and 25.3 in WFCam \textit{J}, \textit{H}, and \textit{K}, respectively\footnote{We use the most recent data releases, DR4 and DR11, for the VIDEO and UKIDSS surveys, respectively.}.  

We assemble our sample of z$\simeq$6.63--6.83 galaxies by applying the above colour cuts to sources identified by running \textsc{SExtractor} \citep{Bertin1996} on a \textit{yYJHK$_s$} and \textit{yYJHK} $\chi^2$ detection image \citep{Szalay1999} over COSMOS and XMM1, respectively. We match the VIDEO and UKIDSS mosaics to the Gaia frame to bring into astrometric agreement with the public HSCSSP and UltraVISTA mosaics. We calculate the optical and near-infrared photometry of each source in 1.2\arcsec{} diameter apertures ($\approx$1.5 times the seeing) and apply aperture corrections determined using the median curve of growth of isolated, unsaturated stars near each source. Photometric errors are calculated as the standard deviation of flux within apertures randomly placed in empty locations (determined using \textsc{SExtractor}) around each source.

To ensure that each source in the sample is real, we require a $>$3$\sigma$ detection in \textit{y}, \textit{Y}, and \textit{J} as well as a $>$5$\sigma$ detection in at least one of those three bands. The requirement of a detection in both the HSC and VIRCam/WFCam imaging removes cross-talk artifacts (see e.g. \citealt{Bowler2017}) from our sample. After applying these selection criteria, our sample consists of 115 sources across COSMOS and 27 across XMM1. There are significantly more sources across COSMOS due to the $\sim$2$\times$ larger area as well as the $\sim$0.5--1 mag increased depth in \textit{y} and \textit{Y}. We also require that each source be undetected ($<$2$\sigma$) in the HSC \textit{g} and \textit{r} bands as they both lie blueward of the Lyman-continuum limit at z$\geq$6.6. These criteria remove 40 and 13 sources in the COSMOS and XMM1 samples, respectively. Furthermore, we remove any T-type brown dwarfs (which can reproduce the above colour cuts) by requiring \textit{Y}-\textit{J}$<$0.45 or (\textit{J}-\textit{H}$>$0 and \textit{J}-\textit{K$_s$}$>$0). These cuts were chosen by simulating colours using brown dwarf templates from the SPEX library \citep{Burgasser2014_SPEX} which show that T-dwarfs have red \textit{Y}-\textit{J} colours yet blue \textit{J}-\textit{H} and \textit{J}-\textit{K$_s$} colours. We emphasize this cut is designed to retain red z$\simeq$6.63--6.83 sources and, as such, mitigates any bias to our inferred [OIII]$+$H$\beta$ EW distribution imposed by preferentially selecting young, blue sources. Applying the T-dwarf cut criterion removes eight sources across COSMOS and one across XMM1. 

Because our goal is to constrain the rest-optical properties of z$\simeq$7 galaxies, we impose magnitude cuts designed to match the \Spitzer{}/IRAC depths. As described below, these depths are typically m$\sim$25--25.5 (3$\sigma$) across COSMOS and XMM1. Consequently, we limit our ground-based sample to similarly luminous sources\footnote{As described in \S\ref{sec:IRAC_Selection}, we also impose a cut based on the IRAC flux uncertainties to ensure we can sufficiently well measure the colours.}, specifically adopting the cut \textit{J}$<$25.7 (\Muv{} $\lesssim$ $-$21.25). To avoid biasing our sample against red (yet still UV bright) sources, we also select those with \textit{K$_s$}$<$25.5. 
These cuts remove 23 sources across COSMOS. All XMM1 sources already satisfy these criteria, largely due to shallower \textit{y} and \textit{Y} depths as noted above. Finally, we visually inspect the HSC, VIRCam, WFCam, and $\chi^2$ detection images of each source satisfying these criteria, removing those due to noise, artificial features, or diffraction spikes which include 3 and 4 sources over COSMOS and XMM1, respectively. 

\defcitealias{Bowler2014}{B14}
\defcitealias{Smit2015}{S15}

\begin{table*}
\centering
\caption{z$\simeq$6.63--6.83 candidates identified across the $\approx$1.5 deg$^2$ and 0.8 deg$^2$ COSMOS and XMM1 fields, respectively. We report \Spitzer{}/IRAC 3.6$\mu$m and 4.5$\mu$m photometry for sources free of strong confusion (see \S\ref{sec:IRAC_Selection}). Sources with an asterisk at the end of their listed ID satisfy additional IRAC criteria and are used to infer the [OIII]$+$H$\beta$ EW distribution at z$\simeq$7 in \S\ref{sec:analysis}. The errors on the [3.6]$-$[4.5] colours reflect simulated 68\% confidence intervals after adding noise to the flux of each source 10,000 times. For sources with a non-detection (S/N$<$1) in one of the IRAC bands, we report the 2$\sigma$ limiting magnitude and colour. We note sources previously identified by \citet[][abbreviated \citetalias{Bowler2014}]{Bowler2014}.} \label{tab:COSMOSXMM1}
\begin{threeparttable}[t]
\begin{tabular}{P{2.0cm}P{1.3cm}P{1.45cm}P{1.3cm}P{1.3cm}P{1.3cm}P{1.3cm}P{1.2cm}P{2.85cm}}
\hline
\hline
Source ID & RA & Dec & \textit{J}\tnote{1} & $\beta_{\mathit{YJHK_s}}$\tnote{2} & 3.6 $\mu$m & 4.5 $\mu$m & [3.6]$-$[4.5] & Notes \Tstrut\Bstrut \\[2pt]
\hline
COS-83688$^{\ast}$ & 09:58:49.21 & +01:39:09.55 & 25.56$^{+0.22}_{-0.18}$ & $-$2.00$\pm$0.57 & 24.47$^{+0.22}_{-0.18}$ & 25.42$^{+0.55}_{-0.36}$ & $-$0.95$^{+0.43}_{-0.58}$ & \Tstrut{} \\[4pt]
COS-87259$^{\ast}$ & 09:58:58.27 & +01:39:20.19 & 24.98$^{+0.15}_{-0.13}$ & $-$0.59$\pm$0.27 & 22.80$^{+0.04}_{-0.04}$ & 22.79$^{+0.05}_{-0.05}$ & 0.01$^{+0.06}_{-0.06}$ & \\[4pt]
COS-160072 & 09:58:54.96 & +01:42:56.68 & 25.55$^{+0.17}_{-0.15}$ & $-$2.98$\pm$0.39 & 25.37$^{+0.62}_{-0.40}$ & 25.68$^{+0.97}_{-0.50}$ & $-$0.31$^{+0.80}_{-1.04}$ & \\[4pt]
COS-237729$^{\ast}$ & 10:00:31.42 & +01:46:51.01 & 25.68$^{+0.18}_{-0.15}$ & $-$1.94$\pm$0.34 & 24.94$^{+0.25}_{-0.20}$ & 25.30$^{+0.34}_{-0.26}$ & $-$0.37$^{+0.36}_{-0.39}$ & \\[4pt]
COS-301652 & 10:00:54.82 & +01:50:05.18 & 25.65$^{+0.22}_{-0.19}$ & $-$2.05$\pm$0.13 & - & - & - & \\[4pt]
COS-312533$^{\ast}$ & 10:00:35.52 & +01:50:38.59 & 25.60$^{+0.17}_{-0.14}$ & $-$1.44$\pm$0.18 & 24.78$^{+0.20}_{-0.17}$ & 25.07$^{+0.28}_{-0.22}$ & $-$0.29$^{+0.30}_{-0.33}$ & \\[4pt]
COS-340502 & 09:59:15.36 & +01:52:00.62 & 25.68$^{+0.20}_{-0.17}$ & $-$2.10$\pm$0.63 & 25.28$^{+0.42}_{-0.30}$ & $>$25.38 & $<$~$-$0.10 & \\[4pt]
COS-369353 & 10:01:59.06 & +01:53:27.75 & 25.59$^{+0.15}_{-0.13}$ & $-$1.54$\pm$0.09 & 23.99$^{+0.10}_{-0.09}$ & 24.59$^{+0.29}_{-0.23}$ & $-$0.60$^{+0.25}_{-0.30}$ & \\[4pt]
COS-378785 & 09:57:22.16 & +01:53:55.50 & 25.38$^{+0.13}_{-0.12}$ & $-$2.21$\pm$0.47 & 23.98$^{+0.61}_{-0.39}$ & 24.35$^{+0.38}_{-0.28}$ & $-$0.36$^{+0.66}_{-0.55}$ & \textit{Y}$-$NB118 = 0.83$^{+0.28}_{-0.33}$ \\[4pt]
COS-400019$^{\ast}$ & 09:59:17.26 & +01:55:03.08 & 25.44$^{+0.14}_{-0.13}$ & $-$2.33$\pm$0.51 & 24.97$^{+0.46}_{-0.32}$ & 25.99$^{+1.70}_{-0.63}$ & $-$1.02$^{+0.79}_{-1.75}$ & \\[4pt]
COS-469110$^{\ast}$ & 10:00:04.36 & +01:58:35.53 & 24.95$^{+0.28}_{-0.22}$ & $-$1.77$\pm$0.55 & 24.20$^{+0.12}_{-0.11}$ & 24.56$^{+0.18}_{-0.16}$ & $-$0.37$^{+0.20}_{-0.21}$ & \\[4pt]
COS-486435 & 10:01:58.71 & +01:59:31.03 & 25.99$^{+0.22}_{-0.19}$ & $-$0.91$\pm$0.24 & 24.56$^{+0.14}_{-0.12}$ & 25.09$^{+0.38}_{-0.28}$ & $-$0.53$^{+0.31}_{-0.40}$ & \\[4pt]
COS-505871$^{\ast}$ & 10:00:21.35 & +02:00:30.93 & 25.51$^{+0.15}_{-0.14}$ & $-$2.23$\pm$0.63 & 24.23$^{+0.14}_{-0.12}$ & 24.34$^{+0.15}_{-0.13}$ & $-$0.11$^{+0.19}_{-0.20}$ & \\[4pt]
COS-534584 & 10:00:42.13 & +02:01:56.87 & 24.99$^{+0.11}_{-0.10}$ & $-$1.77$\pm$0.28 & - & - & - & ID 104600 in \citetalias{Bowler2014} \\[4pt]
COS-559979 & 10:00:42.73 & +02:03:15.30 & 25.98$^{+0.23}_{-0.19}$ & $-$2.28$\pm$0.78 & 26.01$^{+0.85}_{-0.47}$ & $>$25.81 & $<$0.20 & \\[4pt]
COS-593796 & 10:01:53.45 & +02:04:59.62 & 25.57$^{+0.18}_{-0.15}$ & $-$2.68$\pm$0.34 & - & - & - & ID 122368 in \citetalias{Bowler2014} \\[4pt]
COS-596621 & 10:02:07.01 & +02:05:10.18 & 25.59$^{+0.18}_{-0.16}$ & $-$1.87$\pm$0.32 & - & - & - & \textit{Y}$-$NB118 = 0.76$^{+0.34}_{-0.40}$ \\[4pt]
COS-597997 & 09:57:37.00 & +02:05:11.32 & 25.51$^{+0.17}_{-0.15}$ & $-$1.59$\pm$0.25 & 24.42$^{+0.19}_{-0.16}$ & 25.43$^{+0.83}_{-0.46}$ & $-$1.01$^{+0.50}_{-0.83}$ & \textit{Y}$-$NB118 = 0.75$^{+0.38}_{-0.51}$ \\[4pt]
COS-627785$^{\ast}$ & 10:02:05.96 & +02:06:46.09 & 25.49$^{+0.17}_{-0.15}$ & $-$1.67$\pm$0.03 & 24.39$^{+0.14}_{-0.12}$ & 25.22$^{+0.38}_{-0.28}$ & $-$0.83$^{+0.32}_{-0.40}$ & \\[4pt]
COS-632123 & 10:02:08.26 & +02:06:59.45 & 25.61$^{+0.20}_{-0.17}$ & $-$2.81$\pm$0.52 & - & - & - & \\[4pt]
COS-637795 & 10:00:23.48 & +02:07:17.87 & 25.58$^{+0.21}_{-0.17}$ & $-$1.91$\pm$0.42 & 24.87$^{+0.29}_{-0.22}$ & 25.24$^{+0.54}_{-0.36}$ & $-$0.37$^{+0.46}_{-0.58}$ & \\[4pt]
COS-703599$^{\ast}$ & 10:00:34.56 & +02:10:38.01 & 25.70$^{+0.20}_{-0.17}$ & $-$2.21$\pm$0.44 & 24.27$^{+0.17}_{-0.14}$ & 24.74$^{+0.21}_{-0.18}$ & $-$0.47$^{+0.25}_{-0.26}$ & \\[4pt]
COS-705154$^{\ast}$ & 10:00:30.81 & +02:10:42.47 & 25.67$^{+0.16}_{-0.14}$ & $-$2.20$\pm$0.29 & 24.94$^{+0.26}_{-0.21}$ & 25.87$^{+0.71}_{-0.43}$ & $-$0.93$^{+0.51}_{-0.74}$ & \\[4pt]
COS-714412 & 10:00:23.40 & +02:11:09.56 & 25.49$^{+0.15}_{-0.14}$ & $-$2.11$\pm$0.30 & - & - & - & \\[4pt]
COS-759861$^{\ast}$ & 10:02:06.48 & +02:13:24.06 & 24.45$^{+0.05}_{-0.05}$ & $-$1.90$\pm$0.15 & 23.79$^{+0.09}_{-0.09}$ & 24.18$^{+0.16}_{-0.14}$ & $-$0.39$^{+0.17}_{-0.18}$ & ID 169850 in \citetalias{Bowler2014} \\[4pt]
COS-788571$^{\ast}$ & 09:59:21.68 & +02:14:53.02 & 25.27$^{+0.11}_{-0.10}$ & $-$2.12$\pm$0.52 & 24.54$^{+0.16}_{-0.14}$ & 25.56$^{+0.35}_{-0.26}$ & $-$1.02$^{+0.31}_{-0.37}$ & \\[4pt]
COS-795090 & 09:57:23.92 & +02:15:13.73 & 25.50$^{+0.19}_{-0.16}$ & $-$2.49$\pm$0.21 & 24.44$^{+0.62}_{-0.39}$ & 24.82$^{+1.01}_{-0.51}$ & $-$0.38$^{+0.80}_{-1.05}$ & \\[4pt]
COS-810120\tnote{3} & 10:00:30.18 & +02:15:59.68 & 25.07$^{+0.09}_{-0.09}$ & $-$1.69$\pm$0.07 & - & - & - & ID 185070 in \citetalias{Bowler2014} \\[4pt]
COS-857605$^{\ast}$ & 09:59:12.35 & +02:18:28.86 & 25.76$^{+0.23}_{-0.19}$ & $-$1.37$\pm$0.28 & 24.58$^{+0.14}_{-0.13}$ & 24.90$^{+0.21}_{-0.17}$ & $-$0.32$^{+0.22}_{-0.24}$ & \\[4pt]
COS-862541$^{\ast}$\tnote{4} & 10:03:05.25 & +02:18:42.75 & 24.45$^{+0.22}_{-0.18}$ & $-$1.93$\pm$0.40 & 23.33$^{+0.09}_{-0.08}$ & 24.65$^{+0.31}_{-0.24}$ & $-$1.33$^{+0.26}_{-0.32}$ & \\[4pt]
COS-909382 & 10:02:23.12 & +02:21:07.80 & 25.50$^{+0.19}_{-0.17}$ & $-$2.11$\pm$0.89 & - & - & - & \\[4pt]
COS-955126$^{\ast}$ & 09:59:23.62 & +02:23:32.73 & 25.38$^{+0.25}_{-0.20}$ & $-$2.47$\pm$0.14 & 24.46$^{+0.30}_{-0.24}$ & 25.92$^{+1.43}_{-0.59}$ & $-$1.46$^{+0.67}_{-1.43}$ & \\[4pt]
COS-1099982$^{\ast}$ & 10:00:23.37 & +02:31:14.80 & 25.46$^{+0.14}_{-0.13}$ & $-$1.84$\pm$0.10 & 24.13$^{+0.04}_{-0.04}$ & 25.29$^{+0.18}_{-0.15}$ & $-$1.16$^{+0.16}_{-0.18}$ & ID 268576 in \citetalias{Bowler2014} \\[4pt]
COS-1101571 & 09:59:22.43 & +02:31:19.52 & 25.51$^{+0.17}_{-0.15}$ & $-$1.84$\pm$0.12 & - & - & - & \\[4pt]
COS-1136216$^{\ast}$ & 10:01:58.50 & +02:33:08.54 & 24.89$^{+0.12}_{-0.11}$ & $-$2.33$\pm$0.31 & 24.17$^{+0.14}_{-0.12}$ & 24.60$^{+0.19}_{-0.16}$ & $-$0.43$^{+0.21}_{-0.22}$ & ID 279127 in \citetalias{Bowler2014} \\[4pt]
COS-1163765 & 09:58:49.68 & +02:34:35.86 & 25.58$^{+0.15}_{-0.13}$ & $-$1.74$\pm$0.47 & 24.90$^{+0.22}_{-0.18}$ & 26.21$^{+3.11}_{-0.72}$ & $-$1.30$^{+0.76}_{-3.25}$ & \\[4pt]
COS-1224137$^{\ast}$ & 10:01:36.85 & +02:37:49.18 & 24.57$^{+0.08}_{-0.08}$ & $-$1.57$\pm$0.19 & 23.89$^{+0.08}_{-0.08}$ & 24.16$^{+0.12}_{-0.11}$ & $-$0.27$^{+0.14}_{-0.14}$ & ID 304384 in \citetalias{Bowler2014} \\[4pt]
COS-1232922 & 09:57:20.87 & +02:38:20.05 & 25.50$^{+0.19}_{-0.16}$ & $-$1.92$\pm$0.34 & - & - & - & \\[4pt]
COS-1235751$^{\ast}$ & 10:00:11.57 & +02:38:29.81 & 25.63$^{+0.22}_{-0.18}$ & $-$1.10$\pm$0.30 & 24.18$^{+0.18}_{-0.16}$ & 24.33$^{+0.17}_{-0.15}$ & $-$0.16$^{+0.23}_{-0.23}$ & \\[4pt]
\hline
\end{tabular}
\begin{tablenotes}
\item[1] Measured VIRCam and WFCam \textit{J}-band magnitudes for COSMOS and XMM1 sources, respectively.
\item[2] Rest-UV slope (F$_{\nu} \propto \lambda^{\beta+2}$) measured from \textit{YJHK$_s$}/\textit{YJHK} photometry in COSMOS/XMM1.
\item[3] Spectroscopically confirmed to lie at z=6.854 \citep{Smit2018}.
\item[4] Spectroscopically confirmed to lie at z=6.850 with MMT/Binospec (\S\ref{sec:spec_confirmation}).
\end{tablenotes}
\end{threeparttable}
\end{table*}

\begin{table*}
\setcounter{table}{0}
\centering
\caption{Continued.}
\begin{tabular}{P{2.0cm}P{1.3cm}P{1.45cm}P{1.3cm}P{1.3cm}P{1.3cm}P{1.3cm}P{1.2cm}P{2.85cm}}
\hline
\hline
Source ID & RA & Dec & \textit{J}\tnote{1} & $\beta_{\mathit{YJHK_s}}$\tnote{2} & 3.6 $\mu$m & 4.5 $\mu$m & [3.6]$-$[4.5] & Notes \Tstrut\Bstrut \\[2pt]
\hline
COS-1304254 & 10:02:54.04 & +02:42:11.94 & 24.65$^{+0.27}_{-0.22}$ & $-$1.87$\pm$0.29 & - & - & - & \Tstrut{} \\[4pt]
COS-1387948 & 09:59:19.36 & +02:46:41.40 & 25.19$^{+0.16}_{-0.14}$ & $-$1.51$\pm$0.22 & 23.44$^{+0.10}_{-0.09}$ & 23.08$^{+0.08}_{-0.08}$ & 0.36$^{+0.13}_{-0.12}$ & \\[4pt]
XMM1-88011$^{\ast}$ & 02:19:09.49 & $-$05:23:20.75 & 24.92$^{+0.11}_{-0.10}$ & $-$2.13$\pm$0.07 & 24.39$^{+0.23}_{-0.19}$ & 24.54$^{+0.37}_{-0.27}$ & $-$0.15$^{+0.36}_{-0.41}$ & ID 35314 in \citetalias{Bowler2014} \\[4pt]
XMM1-88152 & 02:19:35.14 & $-$05:23:19.31 & 25.36$^{+0.24}_{-0.20}$ & $-$1.17$\pm$0.11 & 24.24$^{+0.34}_{-0.25}$ & 24.80$^{+0.55}_{-0.36}$ & $-$0.57$^{+0.49}_{-0.60}$ & \\[4pt]
XMM1-118414 & 02:19:15.21 & $-$05:19:22.19 & 25.65$^{+0.23}_{-0.19}$ & $-$1.92$\pm$0.40 & - & - & - & \\[4pt]
XMM1-232713 & 02:19:29.13 & $-$05:04:51.68 & 25.42$^{+0.18}_{-0.16}$ & $-$2.62$\pm$0.25 & 25.54$^{+0.58}_{-0.38}$ & 24.85$^{+0.40}_{-0.29}$ & 0.69$^{+0.65}_{-0.55}$ & \\[4pt]
XMM1-250398 & 02:19:33.94 & $-$05:02:39.24 & 25.49$^{+0.24}_{-0.20}$ & $-$2.31$\pm$0.67 & 24.78$^{+0.59}_{-0.38}$ & 24.48$^{+0.83}_{-0.47}$ & 0.30$^{+0.74}_{-0.91}$ & \\[4pt]
XMM1-313310 & 02:16:36.51 & $-$04:54:50.83 & 24.76$^{+0.11}_{-0.10}$ & $-$1.88$\pm$0.19 & - & - & - & \\[4pt]
XMM1-368448 & 02:17:54.37 & $-$04:47:59.61 & 25.28$^{+0.16}_{-0.14}$ & $-$2.63$\pm$0.29 & 24.62$^{+0.39}_{-0.28}$ & 24.66$^{+0.39}_{-0.29}$ & $-$0.04$^{+0.48}_{-0.48}$ & \\[4pt]
XMM1-418672 & 02:17:22.60 & $-$04:39:36.58 & 25.48$^{+0.26}_{-0.21}$ & $-$2.77$\pm$0.37 & 25.03$^{+0.49}_{-0.34}$ & 25.75$^{+1.67}_{-0.63}$ & $-$0.73$^{+0.79}_{-1.66}$ & \\[4pt]
XMM1-426118 & 02:16:55.17 & $-$04:40:42.60 & 25.40$^{+0.19}_{-0.16}$ & $-$1.12$\pm$0.56 & - & - & - & \\[4pt]
\hline
\end{tabular}
\end{table*}

With these selection criteria, we obtain a sample of 50 z$\simeq$6.63--6.83 candidates over COSMOS and XMM1. We find 9 sources over XMM1 and 41 sources over COSMOS, with 38 of those 41 identified over the UltraVISTA ultra-deep stripes. In Table \ref{tab:COSMOSXMM1}, we report the coordinates, \textit{J}-band magnitudes, and rest-UV slopes (measured by fitting F$_{\nu} \propto \lambda^{\beta+2}$ using \textit{YJHK$_s$}/\textit{YJHK} photometry for COSMOS/XMM1) of each source. Our sample consists of sources with magnitudes ranging from \textit{J} = 24.5 to 26.0 and rest-UV slopes ranging from $\beta$ = $-$0.6 to $-$3.0 with a median $\beta$ = $-$1.9 across the entire sample.

While our sample does contain eight sources previously identified by \citet{Bowler2014}, hereafter \citetalias{Bowler2014}, the majority of our sample is composed of newly identified z$\simeq$6.63--6.83 sources. This difference in our sample is partly due the colour-cut selection adopted here versus the photometric redshift selection adopted in \citetalias{Bowler2014}. However, the sample differences are largely driven by advancements in optical/NIR datasets over the COSMOS and XMM1 fields in the past six years. For COSMOS, \citetalias{Bowler2014} used UltraVISTA DR2 whereas here we use DR4 which incorporates 2$-$6$\times$ larger exposure time in the ultra-deep stripes. Furthermore, we use HSCSSP imaging which affords two advantages. First, HSCSSP covers all 1.5 deg$^2$ of COSMOS versus the 1.0 deg$^2$ covered by CFHTLS used in \citetalias{Bowler2014}, enabling us to select sources over all four of the ultra-deep stripes rather than the three from which \citetalias{Bowler2014} selected. Second, HSCSSP adds deep nb921 and \textit{y}-band data enabling a more refined z$\sim$7 selection. Finally, we also gain three sources (COS-469110, COS-862541, and COS-1304254) by selecting over the deep stripes as \citetalias{Bowler2014} restricted their selection to the ultra-deep regions. In XMM1, we use the UKIDSS DR11 while \citetalias{Bowler2014} used DR10 which was shallower by $\sim$0.3 mags\footnote{\url{https://www.nottingham.ac.uk/astronomy/UDS/data/dr11.html}}.

When inferring the rest-optical line strengths of galaxies in this sample, it is useful to know the redshift distribution of sources selected by our adopted colour cuts. In doing so, we are able to estimate the fraction of sources in our sample that are likely to lie in the desired redshift range of z=6.63--6.83, as well as assess the impact that \Lya{} emission has on this fraction. We therefore model the selection completeness of these galaxies as a function of redshift and magnitude by first taking a model SED with rest-UV slope of $\beta$ = $-$2.0. We normalize the SED to \textit{J}-band magnitudes ranging from \textit{J}=23.5--26.5 in 0.1 mag intervals at redshifts ranging from z=6.0--8.0 in 0.01 unit intervals, convolving the SED with the mean IGM transmission function from \citet{Inoue2014} at each redshift. At each magnitude and redshift bin, we calculate the photometry in each filter from the model SED and apply 10,000 realizations of photometric noise set equal to the typical noise of each filter. Because the UltraVISTA data is split into deep and ultra-deep stripes, we simulate the selection completeness for each depth separately. We then calculate completeness as a function of \textit{J}-band magnitude and redshift as the fraction of realizations satisfying our selection criteria described above. 

\begin{figure}
\includegraphics{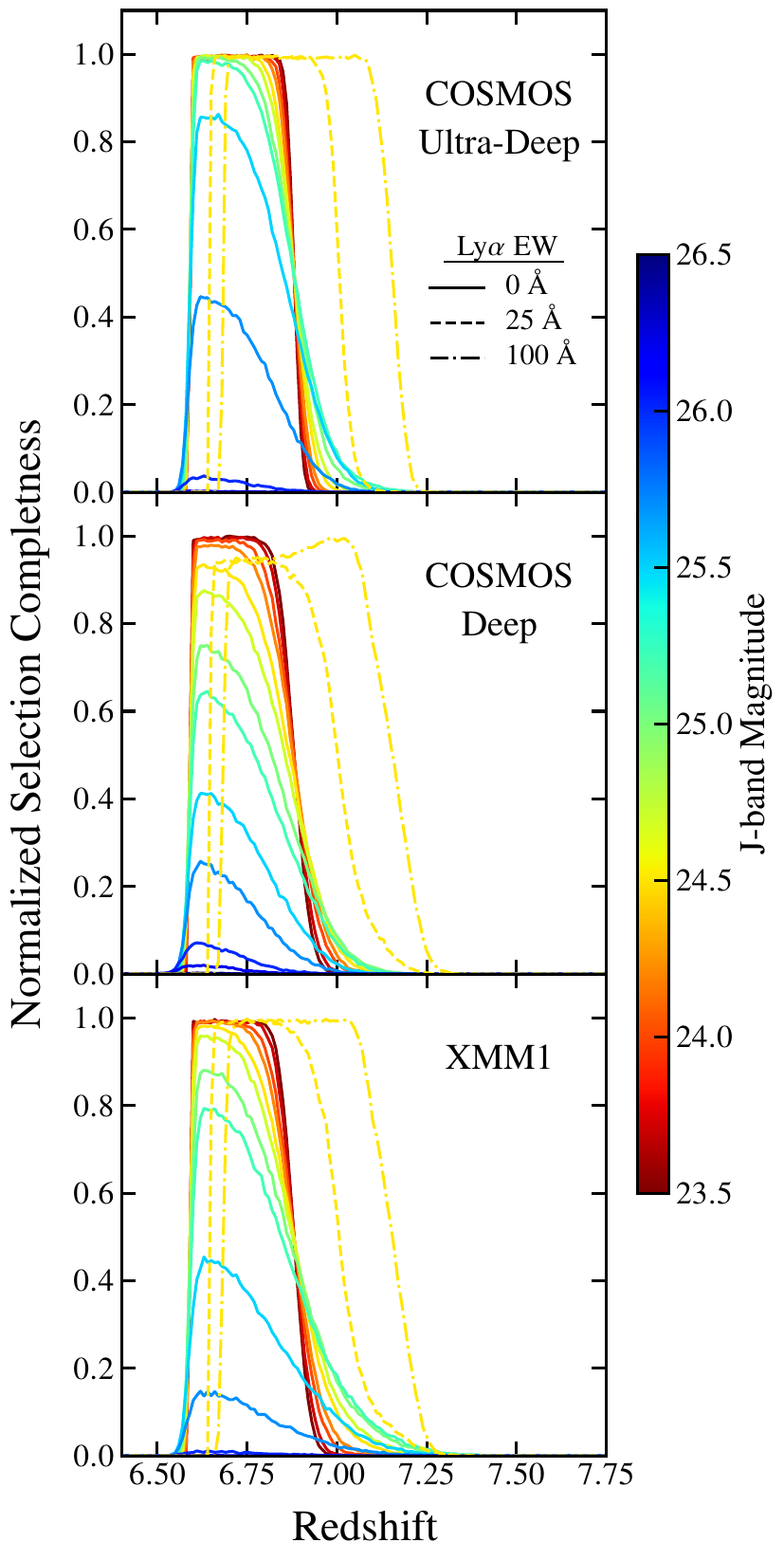}
\caption{Modeled selection completeness as a function of \textit{J}-band magnitude and redshift using typical depths across the UltraVISTA ultra-deep (top) and deep (middle) stripes as well as the XMM1 field (bottom). We show the selection completeness from \textit{J}=23.5--26.5 assuming a \Lya{} EW = 0 \AA{} with solid lines. We also show the selection completeness adopting \Lya{} EW = 25 (dashed) and 100 (dot-dashed) \AA{} for \textit{J}=24.5. Adopting higher \Lya{} EW shifts the selection completeness to higher redshifts (see text).}
\label{fig:Completeness}
\end{figure}

We show our simulated normalized selection completeness for the COSMOS ultra-deep stripes, the COSMOS deep stripes, and XMM1 in Fig. \ref{fig:Completeness}. We present the normalized values because we do not account for the fraction of sources masked by bright foreground sources or artificial features which should be independent of redshift. Using these selection completeness distributions and the \textit{J}-band magnitudes of each source, we estimate that 66\% of the sources in our ground-based sample lie at z=6.63--6.83. We also find that only 13\% are estimated to lie at z$<$6.63 where strong H$\alpha$ would significantly boost the 4.5$\mu$m flux. These estimates are assuming that every source has a \Lya{} EW of 0 \AA{}, again motivated by the rapid decline in \Lya{} emitters observed at z$>$6.5 \citep[e.g.][]{Schenker2014,Pentericci2018}. 

\Lya{} emission pushes our selection completeness towards higher redshifts (see Fig. \ref{fig:Completeness}) because \Lya{} raises the flux in the HSC \textit{y}-band yielding bluer \textit{y}-\textit{Y} colours at fixed redshift. Results from a recent large VLT/FORS2 survey suggest that only $\sim$10\% of \Muv{} $<$ $-$20.25 galaxies at z$\sim$7 exhibit moderate ($>$25 \AA{} EW) \Lya{} emission \citep{Pentericci2018}. If we adopt a conservative assumption that 20\% of the sources in our sample possess \Lya{} emission with 25 \AA{} EW and that the remaining 80\% possess no \Lya{} emission, we infer that 63\% of the sources in our ground-based sample lie at z=6.63--6.83. Therefore, \Lya{} likely has a small impact on the overall fraction of our ground-based sample that lies in the desired redshift window.

\subsubsection{HST-Based Selection over GOODS-N and GOODS-S} \label{sec:GOODS_Selection}

We now describe our selection of 6.6$\lesssim$z$\lesssim$6.9 sources using \HST{} data. Adding these data enables us to push further down the luminosity function and build statistics on the z$\simeq$7 [OIII]$+$H$\beta$ EW distribution. We perform \HST{}-based selection over the GOODS-S and GOODS-N fields because they possess ACS F850LP and WFC3 F105W imaging which offer a valuable probe of the Lyman-break at z$\sim$6--7. The GOODS fields also possess ACS F435W, F606W, and F775W imaging which assist in ruling out low-redshift interlopers. We use the public ACS F435W, F606W, F775W, F814W, and F850LP (hereafter \textit{B$_{435}$}, \textit{V$_{606}$}, \textit{i$_{775}$}, \textit{I$_{814}$}, and \textit{z$_{850}$}, respectively)  as well as the WFC3 F105W, F125W, and F160W (hereafter \textit{Y$_{105}$}, \textit{J$_{125}$}, and \textit{H$_{160}$}, respectively) mosaics from the GOODS \citep{Giavalisco2004} and CANDELS  \citep{Grogin2011,Koekemoer2011} surveys. We astrometrically match the GOODS mosaics to the CANDELS reference frame. 

\defcitealias{Finkelstein2015_LF}{F15}
\defcitealias{Bouwens2015_LF}{B15a}

After simulating \HST{} colours as a function of redshift using the same methods as described above for COSMOS+XMM1, we chose to adopt the following colour cuts over the GOODS fields:
\begin{enumerate}
    \item \textit{z$_{850}$}-\textit{J$_{125}$}$>$1.0
    \item \textit{I$_{814}$}-\textit{J$_{125}$}$>$2.5 or (\textit{I$_{814}$}-\textit{J$_{125}$}$>$1.5 and S/N(\textit{I$_{814}$})$<$2)
    \item \textit{Y$_{105}$}-\textit{J$_{125}$}$<$0.4
\end{enumerate}
Again, fluxes in dropout bands (\textit{I$_{814}$} and \textit{z$_{850}$}) are set to the 1$\sigma$ value in cases of non-detections when applying these criteria. These cuts are largely a more restrictive version of the z$\sim$6.3--7.3 colour cuts presented in \citet{Bouwens2015_LF}, hereafter \citetalias{Bouwens2015_LF}, because we aim to select sources over a narrower redshift interval.

We also require that sources be detected at $>$3$\sigma$ in \textit{Y$_{105}$}, \textit{J$_{125}$}, and \textit{H$_{160}$} (to ensure reasonable WFC3 colour constraints) with a $>$5$\sigma$ detection in at least one of those three bands. After applying these criteria, our \HST{}-based sample consists of 79 sources across GOODS-N and 24 across GOODS-S. Furthermore, we require $<$2$\sigma$ measurements in \textit{B$_{435}$}, \textit{V$_{606}$}, and \textit{i$_{775}$} as common for \HST{}-based z$\sim$7 selection in the literature (e.g. \citealt{McLure2013}; \citetalias{Bouwens2015_LF}). This removes 16 and 2 sources from the GOODS-N and GOODS-S samples, respectively. To mitigate contamination of red low-redshift galaxies, we also enforce \textit{J$_{125}$}-\textit{H$_{160}$}$<$0.3 which corresponds to rest-UV slope cut of $\beta < -0.7$. We expect that this cut will remove very few z$\sim$7 sources because our more luminous COSMOS and XMM1 z$\sim$7 samples include only one source with $\beta > -0.7$ (COS-87259) and because z$\sim$7 galaxies detectable across the GOODS fields have typical rest-UV slopes of $\beta$ $\sim$ $-1.8$ \citep{Bouwens2014_beta}. This \textit{J$_{125}$}-\textit{H$_{160}$} cut removes 11 and 3 sources across GOODS-N and GOODS-S, respectively.

\begin{figure}
\includegraphics{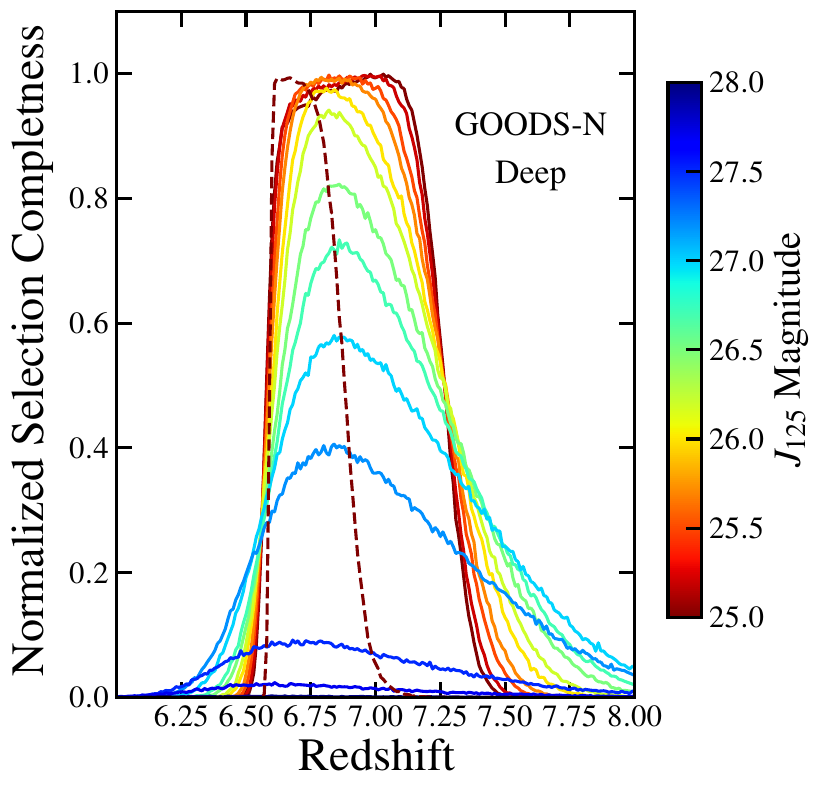}
\caption{Modeled selection completeness as a function of \textit{J$_{125}$}-band magnitude and redshift using typical depths across the CANDELS-Deep GOODS-N region. We show the selection completeness from \textit{J}=25--28 assuming a \Lya{} EW = 0 \AA{} with solid lines. We also show the selection completeness of \textit{J}=25.0 sources over the ultra-deep UltraVISTA region with a dashed line. While the GOODS fields afford the ability to push to fainter magnitudes, the lack of coverage with a second \textit{Y}-band equivalent filter (e.g. WFC3/F098M) leads to a broader redshift selection interval relative to our ground-based selection.}
\label{fig:Completeness_GOODScomparison}
\end{figure}

Due to the lack of coverage with a second Y-band equivalent filter\footnote{The ERS region does have F098M coverage but it lacks F105W coverage. The only region over GOODS with both F098M and F105W coverage is the XDF \citep{Illingworth2013}. However, we ignore the increased depth and F098M coverage over the XDF because it only covers $\sim$4\% of the total GOODS area and most sources within this field would likely be too faint for robust IRAC colour measurements.} (e.g. WFC3/F098M), our \HST{}-based redshift selection is not as precise relative to our ground-based selection. We illustrate this in Fig. \ref{fig:Completeness_GOODScomparison} where we show the modeled selection completeness as a function of redshift and \textit{J$_{125}$} magnitude using the typical depths across the CANDELS-Deep GOODS-N area. In this figure, we also show the modeled selection completeness for \textit{J}=25.0 sources over the ultra-deep UltraVISTA region with a dashed line. It is clear that the GOODS fields allow us to push to much fainter magnitudes. However, the availability of HSC-\textit{z}, HSC-nb921, HSC-\textit{y}, and VIRCam-\textit{Y} photometry enables a much more precise redshift selection over COSMOS and XMM1 relative to GOODS. With these simulated selection completeness distributions and the measured \textit{J$_{125}$} magnitudes of each source, we estimate that 24\% (9\%) of the galaxies in our \HST{}-based sample lie at z=6.63--6.83 (z$<$6.63) assuming \Lya{} EW = 0 \AA{}. None the less, we estimate that 90\% of our GOODS sources lie at z=6.63--7.8 where there is no concern of confusing strong nebular emission with a Balmer break (see below).

We apply the above colour cuts to sources identified by running \textsc{SExtractor} on a \textit{Y$_{105}$J$_{125}$H$_{160}$} $\chi^2$ detection image over both fields. We first calculate the photometry of each source using an elliptical aperture with \citet{Kron1980} parameter of 1.6, consistent with \citetalias{Bouwens2015_LF}. The typical semi-major radius of this aperture for our final GOODS z$\sim$7 sample is 0.50\arcsec{} with typical ellipticity of 1.28. 
For the WFC3 photometry, these measurements are made on mosaics that have been PSF-matched to the \textit{H$_{160}$} band. Measurements in the ACS bands are made on mosaics that have been PSF-matched to the \textit{z$_{850}$} band after reducing the aperture to 70\% of its size from the \textit{Y$_{105}$J$_{125}$H$_{160}$} $\chi^2$ image to maximize signal to noise (see \citetalias{Bouwens2015_LF} for details). We construct the PSF of each band by stacking images of unsaturated isolated stars in each mosaic (at least four in each band), sub-gridding by a factor of ten during stacking as in \citet{Finkelstein2015_LF}, hereafter \citetalias{Finkelstein2015_LF}. To create the PSF convolution kernel for each band, we utilize the Python \textsc{photutils} \citep{Bradley_photutils} package adopting a split cosine bell window function. With this package, we are able to create convolution kernels where the encircled energy distributions agree to within 1\% at radii $>$0.1\arcsec{}. 

We perform aperture flux correction in two steps. First, we multiply by the ratio of flux measured in elliptical apertures with Kron parameters of 2.5 and 1.6 using the square root of the \textit{Y$_{105}$J$_{125}$H$_{160}$} $\chi^2$ image\footnote{The Y$_{105}$ and J$_{125}$ mosaics were PSF-matched to the H$_{160}$ band before we created the $\chi^2$ images.} as in \citetalias{Bouwens2015_LF}. We ignore any sources that have a neighbour within 2.5 Kron parameter elliptical aperture as the flux correction will be overestimated in such instances. Furthermore, because the typical size of the larger aperture (effective radius $\sim$0.6\arcsec{}) is much smaller than the IRAC PSF (FWHM $\sim$ 2\arcsec{}; see below), we cannot reliably deconfuse IRAC photometry for sources with neighbours in such close proximity. This removes 13 and 4 sources from GOODS-N and GOODS-S, respectively. Second, we account for the flux outside the larger aperture using our constructed \textit{H$_{160}$} and \textit{z$_{850}$} PSFs as well as the point-source encircled energy distributions from \citet{Dressel2012} and \citet{Ryon2019} for WFC3 and ACS bands, respectively. We visually inspect all images of each source satisfying the above criteria, verifying that none is due to noise, artificial features, or diffraction spikes. We furthermore verify that each source appears extended, ruling out contamination from brown dwarfs. 

With these selection criteria, we obtain a sample of 15 z$\sim$6.6--7.3 sources over GOODS-S (13 of which fall in the CANDELS-Deep region) and 38 sources over GOODS-N (33 of which fall in the CANDELS-Deep region). We report the coordinates, \textit{J$_{125}$} magnitudes, and rest-UV slopes (measured using \textit{J$_{125}$H$_{160}$} photometry) of each GOODS-N and GOODS-S source in Table \ref{tab:GOODS}. This sample consists of sources with magnitudes ranging from \textit{J$_{125}$} = 25.0 to 27.3 and rest-UV slopes ranging from $\beta$ = $-$0.8 to $-$3.1 with a median $\beta$ = $-$1.7. The biweight mean value is $-$1.78, consistent with that found by \citet{Bouwens2014_beta} for similarly bright z$\sim$7 galaxies. In Table \ref{tab:GOODS}, we note which sources were previously identified as high-redshift galaxies by either \citetalias{Bouwens2015_LF} or \citetalias{Finkelstein2015_LF}. Additionally, \citet{McLure2013} identified GS-897, GS-4218, GS-7199, and GS-19813.

We note that both \citetalias{Bouwens2015_LF} and \citetalias{Finkelstein2015_LF} also report a strong difference in z$\sim$7 source counts between these two fields, with $\approx$2$\times$ the number of z$\sim$7 sources in GOODS-N relative to GOODS-S. While we find a slightly higher ratio ($\approx$2.5), this is to be expected if the GOODS-N (GOODS-S) fields truly represent over(under)-dense regions of the Universe at z$\sim$7 because the z$\sim$7 selection interval we utilize is slightly narrower\footnote{Specifically, our redshift interval is z$\sim$6.6--7.3 while \citetalias{Bouwens2015_LF} and \citetalias{Finkelstein2015_LF} adopt z$\sim$6.3--7.3 and z$\sim$6.5--7.5 intervals, respectively.} than that adopted by both of these previous works. 

\begin{table*}
\centering
\caption{z$\sim$6.6--7.3 candidates identified across the \HST{} GOODS-N and GOODS-S regions. Table format is similar to Table \ref{tab:COSMOSXMM1}. We note sources previously identified as high-redshift galaxies by either \citet{Bouwens2015_LF} or \citet{Finkelstein2015_LF}, abbreviated \citetalias{Bouwens2015_LF} and \citetalias{Finkelstein2015_LF} respectively.} \label{tab:GOODS}
\begin{threeparttable}[t]
\begin{tabular}{P{1.6cm}P{1.3cm}P{1.45cm}P{1.3cm}P{1.3cm}P{1.3cm}P{1.3cm}P{1.2cm}P{3.0cm}}
\hline
\hline
Source ID & RA & Dec & \textit{J$_{125}$}\tnote{1} & $\beta$\tnote{2} & 3.6 $\mu$m & 4.5 $\mu$m & [3.6]$-$[4.5] & Notes \Tstrut\Bstrut \\[2pt]
\hline
GN-2781 & 12:36:29.34 & +62:07:44.84 & 26.82$^{+0.26}_{-0.21}$ & $-$0.87$\pm$1.50 & 25.81$^{+0.40}_{-0.30}$ & 25.64$^{+0.39}_{-0.29}$ & 0.16$^{+0.50}_{-0.49}$ & Identified by \citetalias{Finkelstein2015_LF} \Tstrut{} \\[4pt]
GN-3751$^{\ast}$ & 12:36:22.69 & +62:08:07.96 & 25.24$^{+0.06}_{-0.06}$ & $-$1.97$\pm$0.39 & 25.26$^{+0.18}_{-0.15}$ & 24.73$^{+0.14}_{-0.13}$ & 0.53$^{+0.22}_{-0.21}$ & Identified by \citetalias{Finkelstein2015_LF} \\[4pt]
GN-8480$^{\ast}$ & 12:37:02.92 & +62:09:46.07 & 26.63$^{+0.16}_{-0.14}$ & $-$0.76$\pm$0.94 & 26.03$^{+0.26}_{-0.21}$ & 25.95$^{+0.32}_{-0.25}$ & 0.08$^{+0.36}_{-0.38}$ & \\[4pt]
GN-8853 & 12:35:58.63 & +62:09:51.57 & 25.47$^{+0.25}_{-0.20}$ & $-$1.05$\pm$1.35 & - & - & - & \\[4pt]
GN-15110 & 12:37:00.41 & +62:11:09.38 & 27.11$^{+0.17}_{-0.15}$ & $-$2.88$\pm$1.26 & 26.42$^{+0.32}_{-0.25}$ & $>$26.90 & $<$~$-$0.48 & Identified by \citetalias{Bouwens2015_LF} \& \citetalias{Finkelstein2015_LF} \\[4pt]
GN-15411 & 12:36:52.43 & +62:11:12.74 & 26.65$^{+0.23}_{-0.19}$ & $-$1.44$\pm$1.44 & - & - & - & \\[4pt]
GN-18892 & 12:36:44.01 & +62:17:15.81 & 27.24$^{+0.15}_{-0.14}$ & $-$1.53$\pm$1.05 & 26.62$^{+0.63}_{-0.39}$ & 26.24$^{+0.41}_{-0.30}$ & 0.38$^{+0.69}_{-0.57}$ & Identified by \citetalias{Bouwens2015_LF} \& \citetalias{Finkelstein2015_LF} \\[4pt]
GN-19314$^{\ast}$ & 12:36:37.28 & +62:17:11.27 & 26.40$^{+0.08}_{-0.08}$ & $-$2.75$\pm$0.66 & 25.57$^{+0.17}_{-0.14}$ & 26.30$^{+0.43}_{-0.31}$ & $-$0.73$^{+0.35}_{-0.45}$ & Identified by \citetalias{Bouwens2015_LF} \& \citetalias{Finkelstein2015_LF} \\[4pt]
GN-19764 & 12:36:19.36 & +62:17:13.74 & 26.63$^{+0.24}_{-0.20}$ & $-$2.77$\pm$1.79 & 26.23$^{+0.72}_{-0.42}$ & 26.24$^{+0.62}_{-0.39}$ & $-$0.02$^{+0.80}_{-0.75}$ & \\[4pt]
GN-22985$^{\ast}$ & 12:37:15.73 & +62:16:35.73 & 26.87$^{+0.16}_{-0.14}$ & $-$1.07$\pm$0.88 & 25.71$^{+0.15}_{-0.13}$ & 25.40$^{+0.13}_{-0.12}$ & 0.31$^{+0.19}_{-0.18}$ & Identified by \citetalias{Bouwens2015_LF} \\[4pt]
GN-23141 & 12:36:48.62 & +62:16:31.86 & 26.91$^{+0.17}_{-0.15}$ & $-$2.25$\pm$1.11 & - & - & - & Identified by \citetalias{Bouwens2015_LF} \& \citetalias{Finkelstein2015_LF} \\[4pt]
GN-23420 & 12:36:33.14 & +62:16:27.93 & 26.78$^{+0.17}_{-0.14}$ & $-$1.32$\pm$1.06 & 25.27$^{+0.16}_{-0.14}$ & 25.78$^{+0.33}_{-0.26}$ & $-$0.51$^{+0.30}_{-0.36}$ & Identified by \citetalias{Bouwens2015_LF} \& \citetalias{Finkelstein2015_LF} \\[4pt]
GN-24076 & 12:37:07.83 & +62:16:20.98 & 27.09$^{+0.17}_{-0.15}$ & $-$0.83$\pm$0.93 & - & - & - & Identified by \citetalias{Bouwens2015_LF} \& \citetalias{Finkelstein2015_LF} \\[4pt]
GN-25395 & 12:36:48.91 & +62:16:06.34 & 27.19$^{+0.17}_{-0.15}$ & $-$3.05$\pm$1.44 & 27.08$^{+0.83}_{-0.47}$ & 27.08$^{+0.75}_{-0.44}$ & 0.00$^{+0.93}_{-0.88}$ & Identified by \citetalias{Bouwens2015_LF} \& \citetalias{Finkelstein2015_LF} \\[4pt]
GN-26244$^{\ast}$ & 12:37:03.58 & +62:15:56.86 & 26.42$^{+0.08}_{-0.07}$ & $-$2.24$\pm$0.52 & 25.32$^{+0.13}_{-0.12}$ & 25.93$^{+0.38}_{-0.28}$ & $-$0.62$^{+0.31}_{-0.39}$ & Identified by \citetalias{Bouwens2015_LF} \& \citetalias{Finkelstein2015_LF} \\[4pt]
GN-27138 & 12:36:38.92 & +62:15:49.72 & 26.31$^{+0.14}_{-0.13}$ & $-$2.67$\pm$1.07 & - & - & - & Identified by \citetalias{Bouwens2015_LF} \& \citetalias{Finkelstein2015_LF} \\[4pt]
GN-28450$^{\ast}$ & 12:36:36.48 & +62:15:34.69 & 26.78$^{+0.14}_{-0.12}$ & $-$1.36$\pm$0.85 & 25.07$^{+0.07}_{-0.07}$ & 25.03$^{+0.09}_{-0.08}$ & 0.04$^{+0.11}_{-0.11}$ & Identified by \citetalias{Bouwens2015_LF} \& \citetalias{Finkelstein2015_LF} \\[4pt]
GN-29429$^{\ast}$ & 12:36:19.17 & +62:15:23.29 & 26.05$^{+0.06}_{-0.06}$ & $-$1.58$\pm$0.42 & 24.46$^{+0.07}_{-0.07}$ & 24.62$^{+0.10}_{-0.09}$ & $-$0.16$^{+0.11}_{-0.12}$ & Identified by \citetalias{Bouwens2015_LF} \& \citetalias{Finkelstein2015_LF} \\[4pt]
GN-30721$^{\ast}$ & 12:36:19.84 & +62:15:08.89 & 25.70$^{+0.06}_{-0.05}$ & $-$1.42$\pm$0.37 & 24.36$^{+0.06}_{-0.06}$ & 25.01$^{+0.15}_{-0.13}$ & $-$0.65$^{+0.15}_{-0.16}$ & Identified by \citetalias{Bouwens2015_LF} \& \citetalias{Finkelstein2015_LF} \\[4pt]
GN-33116 & 12:36:49.26 & +62:14:43.92 & 27.26$^{+0.20}_{-0.17}$ & $-$2.82$\pm$1.37 & 26.64$^{+0.62}_{-0.40}$ & 24.42$^{+0.06}_{-0.05}$ & 2.22$^{+0.63}_{-0.40}$ & Identified by \citetalias{Bouwens2015_LF} \& \citetalias{Finkelstein2015_LF} \\[4pt]
GN-33483$^{\ast}$ & 12:36:32.09 & +62:14:40.58 & 26.77$^{+0.16}_{-0.14}$ & $-$2.03$\pm$1.18 & 25.70$^{+0.19}_{-0.16}$ & 25.67$^{+0.17}_{-0.14}$ & 0.03$^{+0.24}_{-0.23}$ & Identified by \citetalias{Bouwens2015_LF} \& \citetalias{Finkelstein2015_LF} \\[4pt]
GN-39066 & 12:36:48.25 & +62:13:38.78 & 26.95$^{+0.19}_{-0.16}$ & $-$2.43$\pm$1.42 & - & - & - & Identified by \citetalias{Bouwens2015_LF} \& \citetalias{Finkelstein2015_LF} \\[4pt]
GN-42769 & 12:36:07.80 & +62:12:59.09 & 25.81$^{+0.09}_{-0.09}$ & $-$1.70$\pm$0.60 & - & - & - & Identified by \citetalias{Bouwens2015_LF} \& \citetalias{Finkelstein2015_LF} \\[4pt]
GN-43467 & 12:37:40.83 & +62:12:52.39 & 26.72$^{+0.22}_{-0.18}$ & $-$2.60$\pm$1.65 & - & - & - & \\[4pt]
GN-47106 & 12:36:54.93 & +62:12:14.43 & 26.19$^{+0.06}_{-0.06}$ & $-$1.52$\pm$0.40 & - & - & - & Identified by \citetalias{Bouwens2015_LF} \& \citetalias{Finkelstein2015_LF} \\[4pt]
GN-47978 & 12:37:20.94 & +62:12:05.72 & 26.64$^{+0.10}_{-0.09}$ & $-$2.01$\pm$0.77 & 25.72$^{+0.20}_{-0.17}$ & 26.03$^{+0.38}_{-0.28}$ & $-$0.31$^{+0.35}_{-0.42}$ & Identified by \citetalias{Bouwens2015_LF} \& \citetalias{Finkelstein2015_LF} \\[4pt]
GN-55562 & 12:36:42.74 & +62:19:27.53 & 26.75$^{+0.11}_{-0.10}$ & $-$3.01$\pm$0.99 & 25.16$^{+0.16}_{-0.14}$ & 26.08$^{+0.42}_{-0.30}$ & $-$0.93$^{+0.34}_{-0.44}$ & Identified by \citetalias{Bouwens2015_LF} \& \citetalias{Finkelstein2015_LF} \\[4pt]
GN-57359 & 12:36:44.97 & +62:19:07.73 & 26.56$^{+0.18}_{-0.15}$ & $-$0.81$\pm$1.02 & 26.04$^{+0.21}_{-0.18}$ & $>$26.77 & $<$~$-$0.72 & Identified by \citetalias{Bouwens2015_LF} \& \citetalias{Finkelstein2015_LF} \\[4pt]
GN-59366$^{\ast}$ & 12:36:56.61 & +62:18:45.00 & 26.27$^{+0.09}_{-0.09}$ & $-$3.09$\pm$0.81 & 25.59$^{+0.15}_{-0.13}$ & 26.13$^{+0.33}_{-0.25}$ & $-$0.55$^{+0.30}_{-0.35}$ & Identified by \citetalias{Finkelstein2015_LF} \\[4pt]
GN-61505$^{\ast}$ & 12:36:52.27 & +62:18:25.50 & 26.76$^{+0.16}_{-0.14}$ & $-$1.04$\pm$0.93 & 26.37$^{+0.29}_{-0.23}$ & 26.37$^{+0.35}_{-0.27}$ & 0.00$^{+0.39}_{-0.42}$ & \\[4pt]
GN-61519 & 12:36:33.58 & +62:18:30.19 & 26.40$^{+0.09}_{-0.08}$ & $-$2.10$\pm$0.65 & 26.02$^{+0.29}_{-0.23}$ & 26.70$^{+0.77}_{-0.44}$ & $-$0.68$^{+0.54}_{-0.79}$ & Identified by \citetalias{Bouwens2015_LF} \& \citetalias{Finkelstein2015_LF} \\[4pt]
GN-61545$^{\ast}$ & 12:36:32.27 & +62:18:28.36 & 26.29$^{+0.08}_{-0.07}$ & $-$2.08$\pm$0.60 & 25.19$^{+0.12}_{-0.10}$ & 25.98$^{+0.33}_{-0.25}$ & $-$0.79$^{+0.28}_{-0.35}$ &  Identified by \citetalias{Bouwens2015_LF} \& \citetalias{Finkelstein2015_LF}\\[4pt]
GN-63360 & 12:36:33.76 & +62:18:08.49 & 27.22$^{+0.25}_{-0.20}$ & $-$1.29$\pm$1.41 & 25.73$^{+0.22}_{-0.19}$ & 25.85$^{+0.27}_{-0.22}$ & $-$0.12$^{+0.31}_{-0.33}$ & Identified by \citetalias{Bouwens2015_LF} \& \citetalias{Finkelstein2015_LF} \\[4pt]
GN-63408$^{\ast}$ & 12:36:37.91 & +62:18:08.59 & 25.66$^{+0.04}_{-0.04}$ & $-$1.34$\pm$0.26 & 25.20$^{+0.13}_{-0.11}$ & 24.37$^{+0.06}_{-0.06}$ & 0.83$^{+0.14}_{-0.13}$ & Identified by \citetalias{Finkelstein2015_LF} \\[4pt]
GN-64262$^{\ast}$ & 12:36:39.61 & +62:18:00.64 & 26.35$^{+0.08}_{-0.07}$ & $-$2.62$\pm$0.60 & 25.45$^{+0.13}_{-0.11}$ & 25.80$^{+0.17}_{-0.15}$ & $-$0.35$^{+0.20}_{-0.21}$ & Identified by \citetalias{Bouwens2015_LF} \& \citetalias{Finkelstein2015_LF} \\[4pt]
GN-66168$^{\ast}$ & 12:37:25.66 & +62:17:43.16 & 24.96$^{+0.05}_{-0.05}$ & $-$1.82$\pm$0.31 & 24.14$^{+0.03}_{-0.03}$ & 24.40$^{+0.05}_{-0.05}$ & $-$0.26$^{+0.06}_{-0.06}$ & Identified by \citetalias{Bouwens2015_LF} \& \citetalias{Finkelstein2015_LF} \\[4pt]
GN-66258 & 12:36:56.45 & +62:17:44.13 & 26.73$^{+0.10}_{-0.09}$ & $-$1.92$\pm$0.72 & $>$26.65 & $>$26.63 & - & Identified by \citetalias{Bouwens2015_LF} \& \citetalias{Finkelstein2015_LF} \\[4pt]
GN-67013 & 12:36:37.11 & +62:17:34.49 & 27.00$^{+0.19}_{-0.16}$ & $-$2.18$\pm$1.11 & 26.62$^{+0.56}_{-0.37}$ & 26.15$^{+0.31}_{-0.24}$ & 0.47$^{+0.61}_{-0.48}$ & Identified by \citetalias{Bouwens2015_LF} \& \citetalias{Finkelstein2015_LF} \\[4pt]
GS-897$^{\ast}$ & 03:32:40.69 & $-$27:44:16.73 & 26.78$^{+0.13}_{-0.12}$ & $-$1.22$\pm$0.90 & 26.96$^{+0.55}_{-0.37}$ & 26.13$^{+0.21}_{-0.18}$ & 0.84$^{+0.58}_{-0.42}$ & Identified by \citetalias{Bouwens2015_LF} \& \citetalias{Finkelstein2015_LF} \\[4pt]
GS-1656$^{\ast}$ & 03:32:36.01 & $-$27:44:41.74 & 26.47$^{+0.09}_{-0.09}$ & $-$1.70$\pm$0.67 & 25.72$^{+0.14}_{-0.13}$ & 25.71$^{+0.17}_{-0.15}$ & 0.01$^{+0.21}_{-0.21}$ & Identified by \citetalias{Bouwens2015_LF} \& \citetalias{Finkelstein2015_LF} \\[4pt]
GS-4218 & 03:32:37.23 & $-$27:45:38.34 & 27.09$^{+0.16}_{-0.14}$ & $-$2.03$\pm$1.06 & 25.99$^{+0.21}_{-0.18}$ & 26.00$^{+0.29}_{-0.23}$ & 0.00$^{+0.31}_{-0.34}$ & Identified by \citetalias{Bouwens2015_LF} \& \citetalias{Finkelstein2015_LF} \\[4pt]
GS-4919 & 03:32:20.01 & $-$27:45:51.79 & 26.87$^{+0.23}_{-0.19}$ & $-$1.18$\pm$1.42 & 26.36$^{+0.71}_{-0.43}$ & 26.36$^{+0.55}_{-0.37}$ & 0.01$^{+0.79}_{-0.70}$ & Identified by \citetalias{Finkelstein2015_LF} \\[4pt]
GS-7199 & 03:32:25.22 & $-$27:46:26.70 & 26.68$^{+0.10}_{-0.10}$ & $-$2.65$\pm$0.91 & 25.73$^{+0.26}_{-0.21}$ & 26.76$^{+0.90}_{-0.49}$ & $-$1.04$^{+0.56}_{-0.93}$ & Identified by \citetalias{Bouwens2015_LF} \& \citetalias{Finkelstein2015_LF} \\[4pt]
\hline
\end{tabular}
\begin{tablenotes}
\item[1] Measured WFC3 F125W magnitudes
\item[2] Rest-UV slope measured from WFC3 F125W and F160W photometry.
\end{tablenotes}
\end{threeparttable}
\end{table*}

\begin{table*}
\setcounter{table}{1}
\centering
\caption{Continued.}
\begin{tabular}{P{2.0cm}P{1.3cm}P{1.45cm}P{1.3cm}P{1.3cm}P{1.3cm}P{1.3cm}P{1.2cm}P{2.85cm}}
\hline
\hline
Source ID & RA & Dec & \textit{J$_{125}$}\tnote{1} & $\beta$\tnote{2} & 3.6 $\mu$m & 4.5 $\mu$m & [3.6]$-$[4.5] & Notes \Tstrut\Bstrut \\[2pt]
\hline
GS-9100$^{\ast}$ & 03:32:50.48 & $-$27:46:55.93 & 26.11$^{+0.08}_{-0.08}$ & $-$1.42$\pm$0.60 & 25.02$^{+0.09}_{-0.08}$ & 25.82$^{+0.23}_{-0.19}$ & $-$0.80$^{+0.21}_{-0.24}$ & Identified by \citetalias{Finkelstein2015_LF} \Tstrut{} \\[4pt]
GS-14372 & 03:32:24.85 & $-$27:48:08.87 & 26.08$^{+0.18}_{-0.16}$ & $-$1.36$\pm$1.32 & - & - & - & Identified by \citetalias{Bouwens2015_LF} \\[4pt]
GS-18358 & 03:32:15.39 & $-$27:48:56.06 & 27.17$^{+0.22}_{-0.18}$ & $-$1.10$\pm$1.30 & 26.62$^{+0.36}_{-0.27}$ & 26.91$^{+0.80}_{-0.46}$ & $-$0.30$^{+0.58}_{-0.83}$ & Identified by \citetalias{Finkelstein2015_LF} \\[4pt]
GS-18406 & 03:32:37.18 & $-$27:48:56.67 & 25.99$^{+0.07}_{-0.06}$ & $-$1.45$\pm$0.47 & - & - & - & Identified by \citetalias{Bouwens2015_LF} \& \citetalias{Finkelstein2015_LF} \\[4pt]
GS-19319 & 03:32:40.06 & $-$27:49:07.53 & 26.87$^{+0.14}_{-0.12}$ & $-$0.99$\pm$0.87 & - & - & - & \\[4pt]
GS-19813 & 03:32:23.77 & $-$27:49:13.64 & 26.41$^{+0.12}_{-0.11}$ & $-$2.27$\pm$0.94 & - & - & - & Identified by \citetalias{Bouwens2015_LF} \& \citetalias{Finkelstein2015_LF} \\[4pt]
GS-27471$^{\ast}$ & 03:32:25.44 & $-$27:50:53.39 & 26.40$^{+0.08}_{-0.07}$ & $-$1.79$\pm$0.58 & 25.07$^{+0.07}_{-0.07}$ & 26.04$^{+0.34}_{-0.26}$ & $-$0.97$^{+0.27}_{-0.34}$ & Identified by \citetalias{Bouwens2015_LF} \& \citetalias{Finkelstein2015_LF} \\[4pt]
GS-28517$^{\ast}$ & 03:32:35.51 & $-$27:51:09.15 & 26.39$^{+0.13}_{-0.12}$ & $-$2.85$\pm$1.05 & 25.63$^{+0.21}_{-0.17}$ & 25.95$^{+0.23}_{-0.19}$ & $-$0.32$^{+0.28}_{-0.29}$ & Identified by \citetalias{Bouwens2015_LF} \& \citetalias{Finkelstein2015_LF} \\[4pt]
GS-36090$^{\ast}$ & 03:32:58.85 & $-$27:53:12.50 & 26.37$^{+0.18}_{-0.15}$ & $-$0.77$\pm$1.03 & 25.36$^{+0.11}_{-0.10}$ & 25.41$^{+0.22}_{-0.18}$ & $-$0.05$^{+0.21}_{-0.24}$ & Identified by \citetalias{Finkelstein2015_LF} \\[4pt]
GS-42630$^{\ast}$ & 03:32:22.48 & $-$27:55:48.90 & 26.10$^{+0.25}_{-0.21}$ & $-$1.48$\pm$1.60 & 26.12$^{+0.31}_{-0.24}$ & 25.81$^{+0.31}_{-0.24}$ & 0.31$^{+0.39}_{-0.39}$ & Identified by \citetalias{Bouwens2015_LF} \& \citetalias{Finkelstein2015_LF} \\[4pt]
\hline
\end{tabular}
\end{table*}

\subsubsection{Additional Selection with Spitzer/IRAC Photometry} \label{sec:IRAC_Selection}

To infer the rest-optical line properties of our z$\simeq$7 sample, we must utilize \Spitzer{}/IRAC photometry. We generate IRAC 3.6$\mu$m and 4.5$\mu$m mosaics across the COSMOS and XMM1 fields by first background subtracting (using \textsc{SExtractor}) each Level 1 corrected basic calibrated data (cbcd) image on the \Spitzer{} Legacy Archive from each of the following surveys: \Spitzer{} Extended Deep Survey (SEDS; \citealt{Ashby2013}), S-CANDELS \citep{Ashby2015}, Star Formation at 4$<$z$<$6 from the \Spitzer{} Large Area Survey with Hyper Suprime-Cam (SPLASH; \citealt{Steinhardt2014}), \Spitzer{} Matching survey of the UltraVISTA ultra-deep Stripes (SMUVS; \citealt{Ashby2018}), and Completing the Legacy of \Spitzer{}/IRAC over COSMOS (P.I. I. Labb\'e). Mosaics are then generated using \textsc{mopex} \citep{Makovoz2005_MOPEX} following the \Spitzer{} Data Analysis Cookbook\footnote{\url{https://irsa.ipac.caltech.edu/data/SPITZER/docs/dataanalysistools/cookbook/}}, and then astrometrically matched to the Gaia reference frame using the IRAF package \textsc{ccmap}. 

Because of IRAC's much larger PSF relative to the ground-based imaging (FWHM$\sim$2\arcsec{} versus $\sim$0.8\arcsec{}), we measure the IRAC photometry for COSMOS and XMM1 sources in 2.8\arcsec{} diameter apertures (consistent with \citetalias{Bowler2014}) and employ a deconfusion method similar to previous studies (e.g. \citealt{Labbe2010}; \citealt{Labbe2013}; \citetalias{Bouwens2015_LF}). Specifically, we convolve the flux profile of each source with a 2D Gaussian having FWHM equal to the quadrature difference of the IRAC FWHM\footnote{We adopt IRAC FWHM values from the IRAC instrument handbook: \url{https://irsa.ipac.caltech.edu/data/SPITZER/docs/irac/iracinstrumenthandbook/}.} and the median seeing from each band in our $\chi^2$ detection images. The flux profile of each source is determined using the square root of the $\chi^2$ image with source footprint given by the \textsc{SExtractor} segmentation map.

After all convolved sources are fit to the IRAC image with their total fluxes as free parameters, we subtract the best-fitting flux profile of each neighbouring source (within a 20\arcsec{} $\times$ 20\arcsec{} square area centered on the source of interest) before measuring the IRAC photometry. Aperture corrections and photometric errors for IRAC photometry are calculated in a similar manner as the HSC and VIRCam photometry. The typical 3$\sigma$ IRAC depth is m=25.3 and m=25.2 for 3.6$\mu$m and 4.5$\mu$m, respectively, over COSMOS and m=24.8 and 24.7 over XMM1. However, due to variable survey coverage (particularly in COSMOS) and source crowding, the depth for a given source can differ up to $\sim$0.5 mag from these values. To mitigate any bias in our inferred rest-optical line strengths from source confusion, we exclude sources where $>$25\% of the flux within the 2.8\arcsec{} diameter aperture is inferred to come from neighbours in either band. Eleven sources in COSMOS and three in XMM1 are removed by this criterion. In Table \ref{tab:COSMOSXMM1}, we report the IRAC photometry and [3.6]$-$[4.5] colours for sources free of strong confusion. 

To ensure that the IRAC data are sufficiently deep for colour measurements, we impose the cut $f_{J}/e_{[4.5]}>2$ where $f_{J}$ is the measured \textit{J}-band flux and $e_{[4.5]}$ is the 1$\sigma$ flux uncertainty in the 4.5$\mu$m band. This cut is equivalent to enforcing the condition that, if a source were as bright in [4.5] as in \textit{J}, it would be detected at S/N$>$2 in [4.5]. An additional 11 and 5 sources are removed by this cut in COSMOS and XMM1, respectively. We do not impose a direct signal-to-noise cut on the IRAC photometry to avoid biasing our inferred [OIII]$+$H$\beta$ EW distribution. None the less, the typical S/N values are 7.8 and 5.0 in [3.6] and [4.5], respectively, for the galaxies satisfying the above condition.

In the GOODS fields, we adopt a slightly modified approach when calculating IRAC photometry given the advantage of having a very high-resolution image of each source (and its neighbours) from \HST{}. Specifically, we measure IRAC photometry in 2.0\arcsec{} diameter apertures (as in \citetalias{Bouwens2015_LF}) from the public S-CANDELS mosaics after astrometrically matching these mosaics to the F160W image in each field. Aperture corrections are calculated by determining the fraction of flux enclosed within the 2.0\arcsec{} diameter aperture after convolving the $\chi^2$ \HST{} flux profile of each target with the local IRAC PSF. We employ a deconfusion approach largely similar to that for COSMOS and XMM1 with the exception that we convolve the flux profile of each source (from the \textit{Y$_{105}$J$_{125}$H$_{160}$} $\chi^2$ image) with an IRAC PSF constructed using unsaturated stars near each source ($<$3.5\arcmin{} separation). We do so as the IRAC depths over the \HST{} GOODS fields are much deeper, and therefore significantly more prone to confusion, than the wide-area COSMOS and XMM1 fields.

Furthermore, we exclude sources based on the residuals in each band after subtracting the best-fitting flux profile of neighbouring sources as well as the source of interest as described in previous works over \HST{} fields \citep[e.g.][]{Labbe2013,Smit2015,Bouwens2015_LF,deBarros2019}. This criteria excludes 13 of the 53 GOODS sources (25\%) which is in good agreement with the $\sim$30\% exclusion rate quoted by \citet{Smit2015}. As in COSMOS+XMM1, we also require $f_{J}/e_{[4.5]}>2$ which removes an additional 18 GOODS sources. Those which satisfy this condition have a typical S/N of 8.4 and 5.5 in [3.6] and [4.5], respectively.

\begin{figure*}
\includegraphics{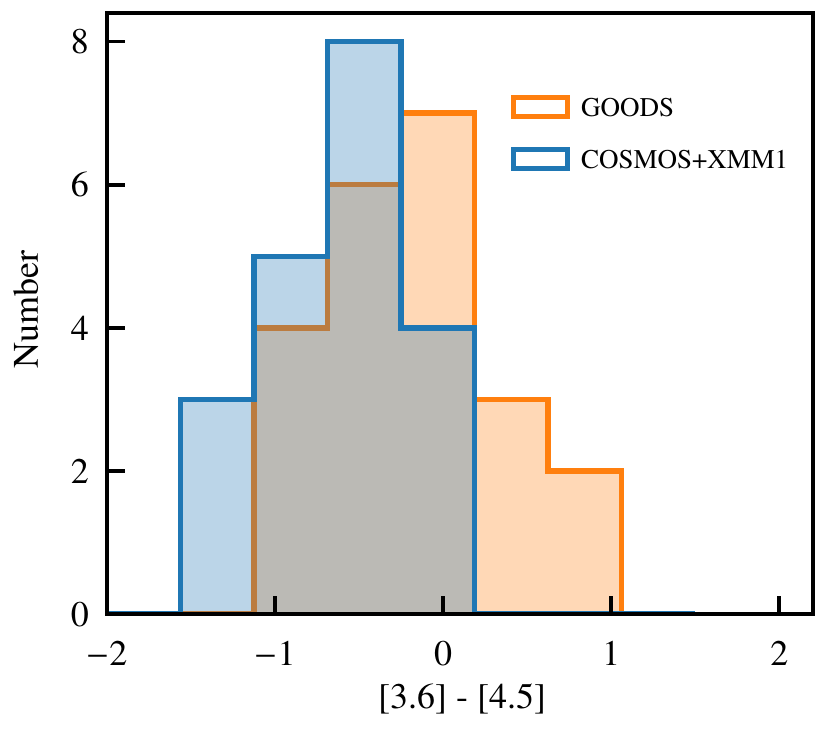}
\includegraphics{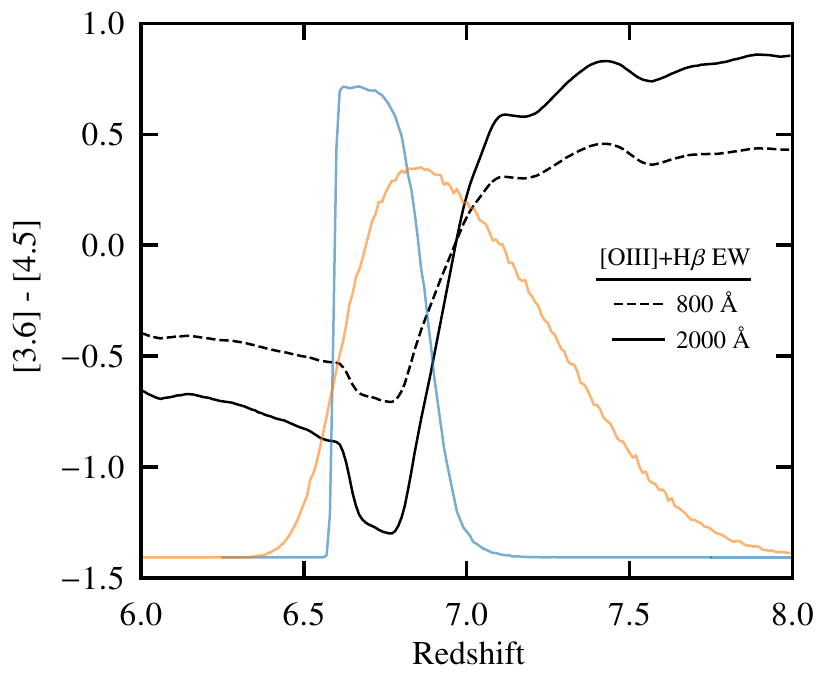}
\caption{
\textbf{Left:} Observed distribution of \Spitzer{}/IRAC [3.6]$-$[4.5] colours in our final z$\simeq$7 sample of 42 sources with robust IRAC measurements. The COSMOS+XMM1 sub-sample is clearly dominated by sources with blue colours as expected for a z$\simeq$6.63--6.83 sample largely comprised of galaxies with strong [OIII]$+$H$\beta$ emission. The IRAC colour distribution for the GOODS sub-sample is notably broader with 13/22 sources having blue colours while the other 9 have red colours. However, these red colours may also be caused by significant [OIII]$+$H$\beta$ emission due to the redshift selection window of this sub-sample and the generally blue rest-UV slopes measured from the \HST{}/WFC3 photometry (see right panel).
\textbf{Right:} Simulated [3.6]$-$[4.5] colours as a function of redshift for galaxies with [OIII]$+$H$\beta$ EW = 800 \AA{} (dashed line) and 2000 \AA{} (solid line). These colours are calculated adopting the empirical and theoretical line ratios described in Appendix \ref{appendix:line_ratios} and a flat rest-optical continuum slope (see \S\ref{sec:analysis}). We also show the simulated normalized selection completeness curves for sources in the ultra-deep COSMOS (\textit{J}=25.0; blue curve) and CANDELS-Deep GOODS-N (\textit{J$_{125}$}=26.5; orange curve) regions assuming \Lya{} EW = 0 \AA{}.  
}
\label{fig:IRACcolourDistn}
\end{figure*}

After applying these additional IRAC criteria, we obtain a final sample of 20 z$\simeq$6.63--6.83 candidates over the COSMOS and XMM1 fields (hereafter referred to as the COSMOS+XMM1 sub-sample) and 22 z$\sim$6.6--7.3 candidates over the GOODS fields (hereafter referred to as the GOODS sub-sample). Each of these sources is denoted with an asterisk at the end of its listed ID in Tables \ref{tab:COSMOSXMM1}$-$\ref{tab:GOODS} and only they are used to infer the z$\simeq$7 [OIII]$+$H$\beta$ EW distribution in \S\ref{sec:analysis}. 

We show the [3.6]$-$[4.5] colour distribution of this final sample in the left panel of Fig. \ref{fig:IRACcolourDistn} for the COSMOS+XMM1 and GOODS samples separately. The COSMOS+XMM1 sub-sample is clearly dominated by sources with blue IRAC colours as expected for a z$\simeq$6.63--6.83 sample largely comprised of galaxies with strong [OIII]$+$H$\beta$ emission. Specifically, we measure blue ([3.6]$-$[4.5] $<$ 0) colours in 19/20 sources and very blue ([3.6]$-$[4.5] $<$ $-$0.8) colours in 8/20 sources with an average measured colour of $-$0.60. The [3.6]$-$[4.5] colour distribution for the GOODS sub-sample is notably broader with 13/22 sources having blue colours while the other 9 have red colours. We emphasize that these red IRAC colours may also be caused by significant [OIII]$+$H$\beta$ emission (see right panel of Fig. \ref{fig:IRACcolourDistn}). This is because 90\% of our GOODS sources are estimated to lie at z=6.63--7.8 (see \S\ref{sec:GOODS_Selection}) and from z$\simeq$7--7.8, [OIII] and H$\beta$ contaminate the 4.5$\mu$m band while the 3.6$\mu$m band probes flux entirely redwards of the Balmer break ($>$3650 \AA{} rest-frame). This is assuming weak dust attenuation which is supported by the generally blue rest-UV slopes measured from the WFC3 photometry.

\subsection{Galaxy Properties} \label{sec:galaxy_properties}

We now infer the galaxy properties, including stellar mass and rest-optical line strengths, of each source in our final sample by fitting their photometry with a photoionization model using the BayEsian Analysis of GaLaxy sEds (BEAGLE; \citealt{Chevallard2016}) code. We use BEAGLE version 0.20.4 which adopts the \citet{Gutkin2016} photoionization models of star-forming galaxies to compute stellar and nebular emission, derived by combining the latest version of \citet{BruzualCharlot2003} stellar population synthesis models into CLOUDY \citep{Ferland2013}. BEAGLE adopts the Bayesian \textsc{multinest} algorithm \citep{Feroz2008,Feroz2009} to compute the posterior probability distribution of galaxy properties from the input photometry.

During this fitting process, we adopt a \citet{Chabrier2003} initial mass function with an upper mass limit of 300 \Msol{}. We consider a delayed star formation history (SFR $\propto \mathrm{t\ e^{-t/\tau}}$) with an allowed age between 1 Myr and the age of the universe at the redshift under consideration, adopting a log-uniform prior on this age. The timescale for exponential decline in the star formation rate is given a log-uniform prior in the range 7 $\leq$ \logten{}($\mathrm{\tau}$/yr) $\leq$ 10.5. We also allow for an ongoing burst of star formation that began 1--10 Myr ago (log-uniform prior) to be superimposed on this delayed star formation history. By permitting ongoing bursts, we allow for the ages implied by blue [3.6]$-$[4.5] colours (indicative of very recent, strong star formation) to differ from the ages implied by a potential Balmer break between the near-IR and 4.5$\mu$m photometry. The sSFR during this burst phase is given a log-uniform prior from 0.1 Gyr$^{-1}$ (the approximate minimum sSFR inferred for massive star-forming galaxies at z$\simeq$2; \citealt{Sanders2018}; \citealt{Strom2018}) to 1000 Gyr$^{-1}$. We also adopt a uniform prior on the redshift between z=0--10 to assess the low-redshift probability of each source.

We adopt an SMC dust prescription \citep{Pei1992} because it well matches the IRX-$\beta$ relation observed at z$\sim$2--3 \citep{Bouwens2016_ALMA,Reddy2018_dust}. The V-band optical depth and metallicity are allowed to vary in the ranges $-$3 $\leq$ \logten{} $\tau_{_V}$ $\leq$ 0.7 and $-$2.2 $\leq$ \logten{}(Z/Z$_{\odot}$) $\leq$ 0.24, respectively, both with log-uniform priors. The dust-to-metal mass ratio and ionization parameter are assigned uniform priors in the ranges 0.1$\leq \xi_d \leq$0.5 and $-$4 $\leq$ \logten{} U$_s$ $\leq$ $-$1, respectively. In this fitting process, we remove \Lya{} from the nebular templates due to the sharp decline in the \Lya{} emitter fraction observed at z$>$6.5 \citep{Schenker2014,Pentericci2018}. It is not currently possible to leave the effective \Lya{} transmission fraction through the intergalactic medium as a free parameter in BEAGLE.

\begin{figure}
\includegraphics{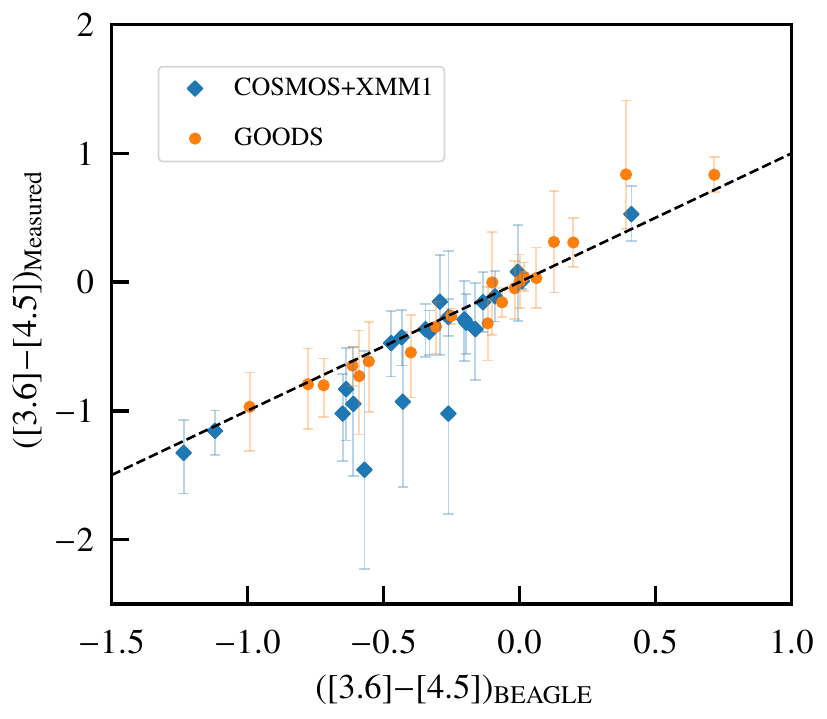}
\caption{Comparison of the measured \Spitzer{}/IRAC [3.6]$-$[4.5] colours versus those inferred by the SED fitting code BEAGLE. Sources from the COSMOS+XMM1 sub-sample are plotted as blue diamonds while those from the GOODS sub-sample are plotted with orange circles. Errorbars are shown on the measured properties (also listed in Tables \ref{tab:COSMOSXMM1} and \ref{tab:GOODS}). We plot the one-to-one relation with a dashed black line which shows that, overall, the modeled colours well reproduce the measured colours.}
\label{fig:BEAGLE_Comparison}
\end{figure}

\begin{figure*}
\includegraphics{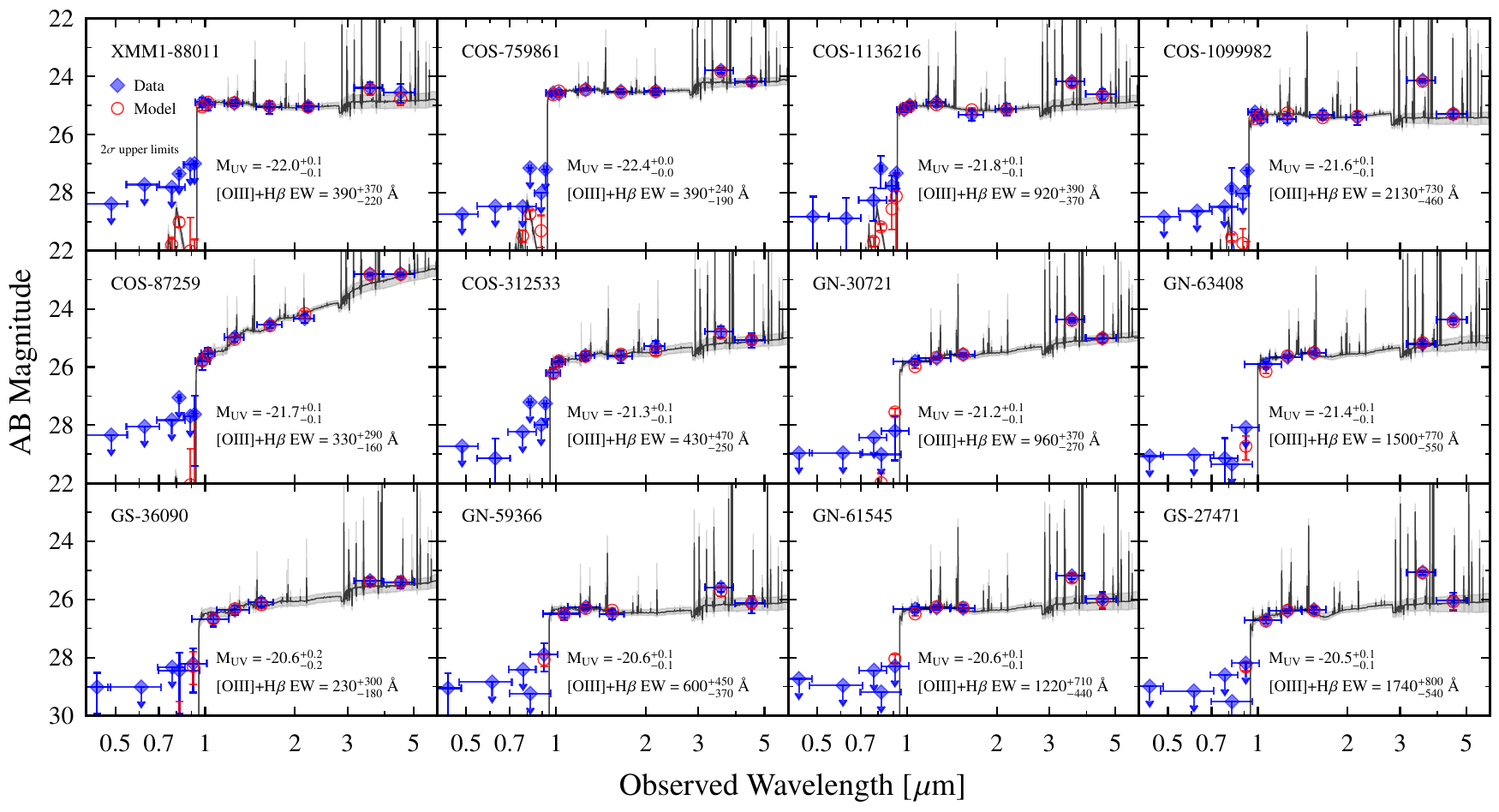}
\caption{Spectral energy distributions of sources in our z$\simeq$7 sample. We select sources to display the variety of inferred [OIII]$+$H$\beta$ EW (generally increasing to the right) and \Muv{} (brighter sources generally towards the top) values. In each panel, the median fit BEAGLE SED model is overlaid in black with gray shading showing the inner 68\% confidence interval from the posterior output by BEAGLE. Blue diamonds show the fitted photometry (with 2$\sigma$ upper limits in cases of non-detections) while the red empty circles show the median model photometry from BEAGLE with errors enclosing the 68\% confidence interval.}
\label{fig:variousSEDs}
\end{figure*}

\begin{figure}
\includegraphics{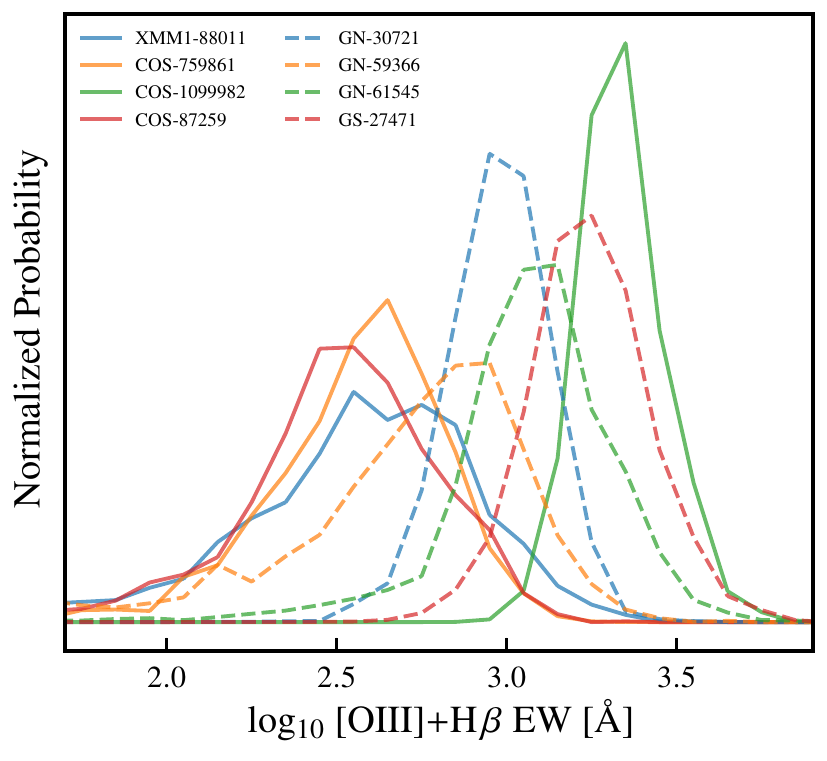}
\caption{Normalized probability distributions of inferred [OIII]$+$H$\beta$ EWs for a subset of sources in our final sample. These distributions were obtained by fitting the optical through mid-infrared photometry of each galaxy with the BEAGLE SED fitting code.}
\label{fig:EWposteriors}
\end{figure}

Before discussing the inferred galaxy properties, we investigate how well the model SEDs from BEAGLE are able to reproduce the measured photometry. To that end, we compare the IRAC [3.6]$-$[4.5] colours as well as the rest-UV slopes calculated from the measured photometry versus those calculated from the model photometry output by BEAGLE\footnote{The model photometry here is taken as the median photometry from the posterior distribution of model SEDs.}. In general, the modeled rest-UV slopes match the observed values within measured uncertainties and furthermore tend to follow a one-to-one relation with the observed values. The IRAC colours output by the BEAGLE models also well-produce the observed colours (see Fig. \ref{fig:BEAGLE_Comparison}), with a maximum offset from the one-to-one relation of 1.2$\sigma$.

We also show, in Fig. \ref{fig:variousSEDs}, the BEAGLE SED fits for various sources in our final sample. The blue diamonds show the observed photometry (with 2$\sigma$ upper limits in cases of non-detections) and the red empty circles show the median model photometry from BEAGLE with errors enclosing the 68\% confidence interval from the posterior. Sources were chosen to display the wide range of inferred [OIII]$+$H$\beta$ EWs (ranging from $\sim$200 to $\gtrsim$2000 \AA{}) and absolute UV magnitudes (ranging from $-$22.5 $\lesssim$ \Muv{} $\lesssim$ $-$20; calculated at 1600 \AA{} rest-frame) across the sample. As illustrated in Fig. \ref{fig:variousSEDs}, the BEAGLE star-forming models are able to reproduce the variety of observed SEDs. 

\begin{table*}
\centering
\caption{Inferred properties of our 42 z$\simeq$7 candidates with robust IRAC photometry over COSMOS, XMM1, GOODS-N, and GOODS-S. These properties were obtained by fitting photometry with a photoionization model using the BEAGLE SED fitting code \citep{Chevallard2016}. The best-fit values and errors are determined by calculating the median and inner 68\% confidence interval values marginalized over the posterior probability distribution function output by BEAGLE.}
\begin{tabular}{P{1.8cm}P{2.5cm}P{1.1cm}P{1.5cm}P{1.3cm}P{2.0cm}P{2.2cm}P{1.8cm}} 
\hline
Source ID & Photometric Redshift & \Muv{} & \logten{}$\left( \Mstar{} / \Msol{} \right)$ & $\tau_{_V}$ & sSFR [Gyr$^{-1}$] & [OIII]+H$\beta$ EW [\AA{}] & Age [Myr] \Tstrut\Bstrut \\
\hline
COS-83688 & 6.70$^{+0.06}_{-0.05}$ & $-$21.6$^{+0.1}_{-0.1}$ & 8.7$^{+0.6}_{-0.5}$ & 0.02$^{+0.05}_{-0.01}$ & 9.3$^{+52.4}_{-8.6}$ & 840$^{+730}_{-500}$ & 37$^{+164}_{-28}$ \Tstrut{} \\[4pt]
COS-87259 & 6.66$^{+0.14}_{-0.06}$ & $-$21.7$^{+0.1}_{-0.1}$ & 10.2$^{+0.6}_{-1.1}$ & 0.53$^{+0.08}_{-0.07}$ & 1.2$^{+9.1}_{-0.9}$ & 330$^{+290}_{-160}$ & 92$^{+433}_{-87}$ \\[4pt]
COS-237729 & 6.87$^{+0.13}_{-0.10}$ & $-$21.1$^{+0.1}_{-0.1}$ & 8.8$^{+0.7}_{-0.7}$ & 0.04$^{+0.10}_{-0.03}$ & 5.0$^{+38.2}_{-4.5}$ & 550$^{+560}_{-360}$ & 37$^{+282}_{-31}$ \\[4pt]
COS-312533 & 6.85$^{+0.08}_{-0.09}$ & $-$21.3$^{+0.1}_{-0.1}$ & 8.9$^{+0.6}_{-0.6}$ & 0.19$^{+0.07}_{-0.10}$ & 6.3$^{+54.4}_{-5.8}$ & 430$^{+470}_{-250}$ & 37$^{+175}_{-30}$ \\[4pt]
COS-400019 & 6.88$^{+0.09}_{-0.10}$ & $-$21.4$^{+0.1}_{-0.1}$ & 8.6$^{+0.5}_{-0.4}$ & 0.01$^{+0.04}_{-0.01}$ & 6.1$^{+49.6}_{-5.5}$ & 450$^{+540}_{-240}$ & 38$^{+131}_{-28}$ \\[4pt]
COS-469110 & 6.63$^{+0.09}_{-0.02}$ & $-$21.6$^{+0.1}_{-0.1}$ & 9.2$^{+0.6}_{-0.8}$ & 0.09$^{+0.09}_{-0.07}$ & 6.6$^{+33.1}_{-5.8}$ & 640$^{+420}_{-330}$ & 62$^{+298}_{-55}$ \\[4pt]
COS-505871 & 6.63$^{+0.11}_{-0.05}$ & $-$21.2$^{+0.1}_{-0.1}$ & 10.2$^{+0.1}_{-1.7}$ & 0.03$^{+0.13}_{-0.02}$ & 0.8$^{+2.3}_{-0.5}$ & 220$^{+350}_{-160}$ & 311$^{+240}_{-305}$ \\[4pt]
COS-627785 & 6.66$^{+0.05}_{-0.03}$ & $-$21.5$^{+0.1}_{-0.1}$ & 8.6$^{+0.6}_{-0.4}$ & 0.03$^{+0.06}_{-0.03}$ & 10.4$^{+57.5}_{-9.8}$ & 1100$^{+720}_{-510}$ & 22$^{+98}_{-16}$ \\[4pt]
COS-703599 & 6.65$^{+0.08}_{-0.04}$ & $-$21.1$^{+0.1}_{-0.1}$ & 8.6$^{+1.1}_{-0.4}$ & 0.07$^{+0.10}_{-0.06}$ & 7.5$^{+62.4}_{-6.6}$ & 870$^{+630}_{-370}$ & 15$^{+413}_{-11}$ \\[4pt]
COS-705154 & 6.69$^{+0.05}_{-0.03}$ & $-$21.4$^{+0.1}_{-0.1}$ & 8.7$^{+0.5}_{-0.4}$ & 0.01$^{+0.03}_{-0.01}$ & 2.9$^{+20.3}_{-2.5}$ & 520$^{+460}_{-300}$ & 42$^{+116}_{-31}$ \\[4pt]
COS-759861 & 6.69$^{+0.03}_{-0.03}$ & $-$22.4$^{+0.0}_{-0.0}$ & 9.7$^{+0.3}_{-0.9}$ & 0.04$^{+0.07}_{-0.03}$ & 2.4$^{+8.8}_{-2.1}$ & 390$^{+240}_{-190}$ & 144$^{+198}_{-130}$ \\[4pt]
COS-788571 & 6.63$^{+0.02}_{-0.02}$ & $-$21.5$^{+0.1}_{-0.1}$ & 8.5$^{+0.4}_{-0.4}$ & 0.01$^{+0.02}_{-0.00}$ & 21.7$^{+76.4}_{-20.8}$ & 1360$^{+910}_{-550}$ & 23$^{+60}_{-18}$ \\[4pt]
COS-857605 & 6.77$^{+0.09}_{-0.08}$ & $-$21.3$^{+0.1}_{-0.1}$ & 9.3$^{+0.4}_{-0.6}$ & 0.23$^{+0.07}_{-0.08}$ & 1.6$^{+8.4}_{-1.3}$ & 310$^{+290}_{-190}$ & 70$^{+199}_{-55}$ \\[4pt]
COS-862541 & 6.72$^{+0.04}_{-0.04}$ & $-$22.5$^{+0.1}_{-0.1}$ & 8.8$^{+0.5}_{-0.3}$ & 0.01$^{+0.03}_{-0.01}$ & 67.6$^{+114.2}_{-53.9}$ & 2250$^{+880}_{-610}$ & 25$^{+192}_{-21}$ \\[4pt]
COS-955126 & 6.70$^{+0.05}_{-0.04}$ & $-$21.6$^{+0.1}_{-0.1}$ & 8.6$^{+0.5}_{-0.4}$ & 0.01$^{+0.02}_{-0.01}$ & 6.8$^{+46.3}_{-6.3}$ & 750$^{+780}_{-440}$ & 34$^{+109}_{-24}$ \\[4pt]
COS-1099982 & 6.67$^{+0.03}_{-0.02}$ & $-$21.6$^{+0.1}_{-0.1}$ & 8.5$^{+0.6}_{-0.3}$ & 0.02$^{+0.04}_{-0.02}$ & 50.8$^{+77.8}_{-42.6}$ & 2130$^{+730}_{-460}$ & 28$^{+211}_{-23}$ \\[4pt]
COS-1136216 & 6.63$^{+0.03}_{-0.01}$ & $-$21.8$^{+0.1}_{-0.1}$ & 8.5$^{+0.7}_{-0.3}$ & 0.01$^{+0.03}_{-0.01}$ & 9.1$^{+61.7}_{-8.5}$ & 920$^{+390}_{-370}$ & 11$^{+183}_{-6}$ \\[4pt]
COS-1224137 & 6.78$^{+0.06}_{-0.06}$ & $-$22.3$^{+0.1}_{-0.1}$ & 9.1$^{+0.7}_{-0.5}$ & 0.17$^{+0.06}_{-0.06}$ & 3.9$^{+28.6}_{-3.5}$ & 370$^{+190}_{-140}$ & 23$^{+140}_{-16}$ \\[4pt]
COS-1235751 & 6.74$^{+0.09}_{-0.07}$ & $-$21.3$^{+0.1}_{-0.1}$ & 9.6$^{+0.4}_{-0.8}$ & 0.33$^{+0.07}_{-0.07}$ & 1.5$^{+7.7}_{-1.2}$ & 300$^{+250}_{-180}$ & 78$^{+229}_{-66}$ \\[4pt]
XMM1-88011 & 6.71$^{+0.07}_{-0.06}$ & $-$22.0$^{+0.1}_{-0.1}$ & 9.0$^{+0.6}_{-0.6}$ & 0.01$^{+0.04}_{-0.01}$ & 2.0$^{+14.5}_{-1.7}$ & 390$^{+370}_{-220}$ & 33$^{+203}_{-26}$ \\[4pt]
GN-3751 & 7.35$^{+0.12}_{-0.13}$ & $-$21.8$^{+0.1}_{-0.1}$ & 8.6$^{+0.6}_{-0.3}$ & 0.01$^{+0.03}_{-0.01}$ & 4.9$^{+39.5}_{-4.4}$ & 780$^{+500}_{-360}$ & 19$^{+79}_{-12}$ \\[4pt]
GN-8480 & 6.96$^{+0.23}_{-0.23}$ & $-$20.4$^{+0.1}_{-0.1}$ & 8.7$^{+0.5}_{-0.8}$ & 0.07$^{+0.11}_{-0.06}$ & 1.7$^{+16.0}_{-1.4}$ & 320$^{+410}_{-210}$ & 63$^{+262}_{-55}$ \\[4pt]
GN-19314 & 6.73$^{+0.07}_{-0.07}$ & $-$20.4$^{+0.1}_{-0.1}$ & 8.1$^{+0.8}_{-0.4}$ & 0.01$^{+0.05}_{-0.01}$ & 5.5$^{+32.1}_{-5.0}$ & 870$^{+570}_{-420}$ & 19$^{+181}_{-13}$ \\[4pt]
GN-22985 & 6.60$^{+0.50}_{-0.21}$ & $-$20.0$^{+0.2}_{-0.2}$ & 8.9$^{+0.6}_{-0.8}$ & 0.32$^{+0.08}_{-0.09}$ & 1.8$^{+17.1}_{-1.5}$ & 270$^{+310}_{-200}$ & 47$^{+217}_{-40}$ \\[4pt]
GN-26244 & 6.80$^{+0.05}_{-0.06}$ & $-$20.5$^{+0.1}_{-0.1}$ & 8.3$^{+1.0}_{-0.4}$ & 0.03$^{+0.09}_{-0.02}$ & 6.5$^{+62.6}_{-6.0}$ & 940$^{+620}_{-480}$ & 11$^{+272}_{-6}$ \\[4pt]
GN-28450 & 6.48$^{+0.10}_{-0.10}$ & $-$20.1$^{+0.1}_{-0.1}$ & 10.0$^{+0.1}_{-1.5}$ & 0.03$^{+0.29}_{-0.02}$ & 0.5$^{+49.1}_{-0.3}$ & 110$^{+790}_{-80}$ & 336$^{+222}_{-218}$ \\[4pt]
GN-29429 & 6.77$^{+0.09}_{-0.10}$ & $-$20.8$^{+0.1}_{-0.1}$ & 10.1$^{+0.2}_{-1.9}$ & 0.06$^{+0.21}_{-0.05}$ & 0.8$^{+32.3}_{-0.5}$ & 190$^{+330}_{-140}$ & 277$^{+217}_{-273}$ \\[4pt]
GN-30721 & 6.78$^{+0.04}_{-0.04}$ & $-$21.2$^{+0.1}_{-0.1}$ & 8.8$^{+0.7}_{-0.4}$ & 0.18$^{+0.06}_{-0.07}$ & 8.6$^{+32.3}_{-7.8}$ & 960$^{+370}_{-270}$ & 21$^{+163}_{-14}$ \\[4pt]
GN-33483 & 6.93$^{+0.31}_{-0.26}$ & $-$20.2$^{+0.2}_{-0.2}$ & 8.9$^{+0.6}_{-0.9}$ & 0.17$^{+0.11}_{-0.13}$ & 1.4$^{+12.1}_{-1.1}$ & 290$^{+360}_{-190}$ & 52$^{+347}_{-45}$ \\[4pt]
GN-59366 & 6.79$^{+0.08}_{-0.07}$ & $-$20.6$^{+0.1}_{-0.1}$ & 8.6$^{+0.7}_{-0.7}$ & 0.01$^{+0.04}_{-0.01}$ & 3.4$^{+23.9}_{-3.0}$ & 600$^{+450}_{-370}$ & 47$^{+269}_{-40}$ \\[4pt]
GN-61505 & 6.68$^{+0.32}_{-0.16}$ & $-$20.2$^{+0.1}_{-0.1}$ & 8.3$^{+0.6}_{-0.6}$ & 0.02$^{+0.07}_{-0.02}$ & 2.4$^{+20.1}_{-2.1}$ & 380$^{+340}_{-240}$ & 41$^{+193}_{-31}$ \\[4pt]
GN-61545 & 6.77$^{+0.04}_{-0.06}$ & $-$20.6$^{+0.1}_{-0.1}$ & 8.2$^{+1.0}_{-0.4}$ & 0.02$^{+0.06}_{-0.02}$ & 12.8$^{+71.0}_{-12.0}$ & 1220$^{+710}_{-440}$ & 13$^{+282}_{-9}$ \\[4pt]
GN-63408 & 7.22$^{+0.13}_{-0.12}$ & $-$21.4$^{+0.1}_{-0.1}$ & 8.5$^{+0.3}_{-0.2}$ & 0.12$^{+0.05}_{-0.07}$ & 13.8$^{+75.8}_{-13.3}$ & 1500$^{+770}_{-550}$ & 8$^{+16}_{-3}$ \\[4pt]
GN-64262 & 6.67$^{+0.09}_{-0.08}$ & $-$20.5$^{+0.1}_{-0.1}$ & 9.1$^{+0.4}_{-1.2}$ & 0.01$^{+0.04}_{-0.01}$ & 2.5$^{+18.0}_{-2.1}$ & 470$^{+450}_{-290}$ & 203$^{+344}_{-196}$ \\[4pt]
GN-66168 & 6.80$^{+0.04}_{-0.04}$ & $-$21.9$^{+0.0}_{-0.0}$ & 8.9$^{+1.0}_{-0.5}$ & 0.12$^{+0.06}_{-0.06}$ & 2.5$^{+16.4}_{-2.0}$ & 370$^{+150}_{-120}$ & 17$^{+336}_{-12}$ \\[4pt]
GS-897 & 7.32$^{+0.19}_{-0.18}$ & $-$20.3$^{+0.1}_{-0.1}$ & 8.0$^{+0.6}_{-0.3}$ & 0.01$^{+0.04}_{-0.01}$ & 7.3$^{+56.9}_{-6.6}$ & 810$^{+670}_{-430}$ & 20$^{+95}_{-14}$ \\[4pt]
GS-1656 & 6.94$^{+0.12}_{-0.12}$ & $-$20.6$^{+0.1}_{-0.1}$ & 9.2$^{+0.3}_{-1.4}$ & 0.03$^{+0.07}_{-0.03}$ & 0.8$^{+4.2}_{-0.6}$ & 220$^{+280}_{-180}$ & 235$^{+295}_{-230}$ \\[4pt]
GS-9100 & 6.80$^{+0.04}_{-0.06}$ & $-$20.8$^{+0.1}_{-0.1}$ & 8.3$^{+0.5}_{-0.3}$ & 0.12$^{+0.08}_{-0.10}$ & 23.5$^{+102.8}_{-22.4}$ & 1320$^{+630}_{-440}$ & 10$^{+72}_{-6}$ \\[4pt]
GS-27471 & 6.76$^{+0.04}_{-0.06}$ & $-$20.5$^{+0.1}_{-0.1}$ & 8.3$^{+0.5}_{-0.3}$ & 0.03$^{+0.09}_{-0.03}$ & 63.9$^{+113.2}_{-58.0}$ & 1740$^{+800}_{-540}$ & 19$^{+253}_{-15}$ \\[4pt]
GS-28517 & 6.84$^{+0.28}_{-0.20}$ & $-$20.5$^{+0.1}_{-0.1}$ & 8.9$^{+0.4}_{-1.1}$ & 0.03$^{+0.08}_{-0.03}$ & 2.2$^{+18.7}_{-1.9}$ & 360$^{+470}_{-260}$ & 122$^{+311}_{-115}$ \\[4pt]
GS-36090 & 6.82$^{+0.24}_{-0.23}$ & $-$20.6$^{+0.2}_{-0.2}$ & 9.3$^{+0.3}_{-0.6}$ & 0.24$^{+0.09}_{-0.11}$ & 1.4$^{+8.0}_{-1.1}$ & 230$^{+300}_{-180}$ & 91$^{+253}_{-74}$ \\[4pt]
GS-42630 & 7.27$^{+0.20}_{-0.19}$ & $-$21.0$^{+0.1}_{-0.1}$ & 8.6$^{+0.4}_{-0.4}$ & 0.01$^{+0.03}_{-0.01}$ & 3.7$^{+23.7}_{-3.2}$ & 380$^{+410}_{-220}$ & 42$^{+96}_{-29}$ \\[4pt]
\hline
\end{tabular}
\label{tab:galaxy_properties}
\end{table*}

We report the inferred photometric redshifts, absolute UV magnitudes, stellar masses, V-band optical depths, sSFRs (over the 1--10 Myr `burst' timescale), [OIII]+H$\beta$ EWs, and ages (time since first star formed) of each source in Table \ref{tab:galaxy_properties}. The best-fit values and errors are determined by calculating the median and inner 68\% confidence interval values marginalized over the posterior probability distribution function output by BEAGLE. All sources in our sample are inferred to lie at z$>$4 with $>$99.95\% probability. The photometric redshifts of the COSMOS+XMM1 and GOODS sub-samples span z=6.62--6.91 and z=6.50--7.64, respectively, consistent with the redshift completeness simulations described in \S\ref{sec:sample_selection}.

For the COSMOS+XMM1 sub-sample, we infer absolute UV magnitudes ranging from $-$22.5 $\leq$ \Muv{} $\leq$ $-$21.1. Inferred stellar masses span the range 8.5 $\leq$ \logten{}$\left(\Mstar{}/\Msol{}\right)$ $\leq$ 10.2, suggesting that this sub-sample contains some of the most luminous (\citetalias{Bouwens2015_LF}; \citetalias{Finkelstein2015_LF}; \citealt{Bowler2017}) and massive \citep{Duncan2014,Song2016} sources present in the Universe at z$\simeq$7. Unsurprisingly, the \HST{}-based GOODS sub-sample is largely composed of less luminous sources with \Muv{} $>$ $-$21. There is more overlap in the stellar mass distribution between the two sub-samples because of significant ($\sim$0.3--0.4 dex) scatter in the \Muv{}$-$\Mstar{} relation \citep{Duncan2014,Salmon2015,Song2016}. None the less, while all sources in the COSMOS+XMM1 sub-sample have inferred stellar masses of \logten{}$\left(\Mstar{}/\Msol{}\right)$ $\geq$ 8.5, only 68\% (15/22) of sources in GOODS are inferred to be so massive. Therefore, the GOODS sub-sample is generally composed of less massive sources as well.

Both the COSMOS+XMM1 and GOODS sub-samples possess sources with inferred [OIII]$+$H$\beta$ EWs ranging from $\sim$200 to $\gtrsim$2000 \AA{}. As expected, those with the largest inferred EWs ($>$1200 \AA{}) show strong IRAC colours ($|[3.6]-[4.5]| \gtrsim 0.8$). These extreme emission line galaxies are inferred to possess very large sSFRs with an average value of 36 Gyr$^{-1}$, much higher than the typical sSFR of our sample (4.4 Gyr$^{-1}$), suggesting that they are undergoing a phase of intense star formation. Our sample also contains sources with flat IRAC colours, suggesting relatively low [OIII]+H$\beta$ emission ($\lesssim$300 \AA{} EW; for example, see COS-87259 and GS-36090 in Fig. \ref{fig:variousSEDs}) and small sSFR (1.2 Gyr$^{-1}$ on average). 

In Fig. \ref{fig:EWposteriors}, we show the posterior [OIII]$+$H$\beta$ EW distributions for a representative set of galaxies across our sample (a subset of those shown in Fig. \ref{fig:variousSEDs}). In general, sources with stronger (i.e. non-flat) IRAC colours have more precise inferred [OIII]$+$H$\beta$ EW posteriors (in logarithmic space). Those with median inferred [OIII]$+$H$\beta$ EWs $<$400 \AA{} have a typical 68\% confidence interval uncertainty of $\pm$0.36 dex on their EW distributions, much larger than the typical $\pm$0.17 dex uncertainty for sources with median inferred [OIII]$+$H$\beta$ EWs $>$1200 \AA{}. This strong anti-correlation of median inferred EW and posterior uncertainty is largely due to the scaling of IRAC colour with EW. A z=6.75 source with flat rest-optical continuum and measured [3.6]$-$[4.5] = 0.3$\pm$0.3 mags has an inferred [OIII]$+$H$\beta$ EW\footnote{We use the line ratios detailed in Appendix \ref{appendix:line_ratios} to calculate these inferred EWs.} of 270$^{\scaleto{+370}{4.5pt}}_{\scaleto{-270}{4.5pt}}$ \AA{} while a source with [3.6]$-$[4.5] = 1.0$\pm$0.3 mags has an inferred EW of 1330$^{\scaleto{+690}{4.5pt}}_{\scaleto{-530}{4.5pt}}$ \AA{}. Therefore, sources with stronger IRAC colours will have smaller inferred EW uncertainties (in dex) at fixed colour measurement uncertainty. The typical inferred [OIII]$+$H$\beta$ EW uncertainty (68\% confidence interval) across all sources in our sample is $\pm$0.33 dex.

We also note that our COSMOS sub-sample contains three sources (COS-378785, COS-596621, and COS-597997) that have a significant excess in the VIRCam NB118 filter with \textit{Y}$-$NB118 $>$ 0.7 (see Table \ref{tab:COSMOSXMM1}). We explore whether this could plausibly be due to strong rest-UV nebular emission. Assuming a redshift range of z$\simeq$6.63--6.83, the NB118 filter (1.18--1.20$\mu$m) covers a rest-frame wavelength of $\approx$1510--1570 \AA{}. Previous spectroscopic studies have discovered strong CIV$\lambda$1548,1550 emission at z$>$6 with measured EWs ranging from $\sim$10--40 \AA{} \citep{Stark2015_CIV,Mainali2017,Schmidt2017}. Because the EWs implied by the \textit{Y}$-$NB118 colours\footnote{We use the formula EW$_{\mathrm{min}}$ = $\left(10^{0.4 \times \left(Y-\mathrm{NB118}\right)} - 1\right) \times \mathrm{BW}/\left(1+z\right)$ to estimate the minimum EW implied by a given \textit{Y}$-$NB118 colour where BW is the bandwidth of the NB118 filter (117 \AA{}). If the emission line lies at an observed wavelength of non-maximum transmission through NB118, the EW will be boosted by a factor T($\lambda$)/T$_{\mathrm{max}}$. We quote EWs using the range T($\lambda$)/T$_{\mathrm{max}}$ = 0.5--1.} ($\sim$15--40 \AA{}) are consistent with these past measurements, we interpret the strong NB118 excesses as tentative signatures of strong CIV emission at z$\approx$6.66--6.72. While such strong CIV emission can be powered by low-metallicity ($\lesssim$0.1 Z$_{\odot}$), young stellar populations ($<$50 Myr; \citealt{Senchyna2019}), it can also arise from AGN activity. Fortunately, rest-UV spectral diagnostics can be used to determine the primary mechanism powering this emission \citep[e.g.][]{Feltre2016,Mainali2017,Hirschmann2019}.

\begin{figure}
\includegraphics{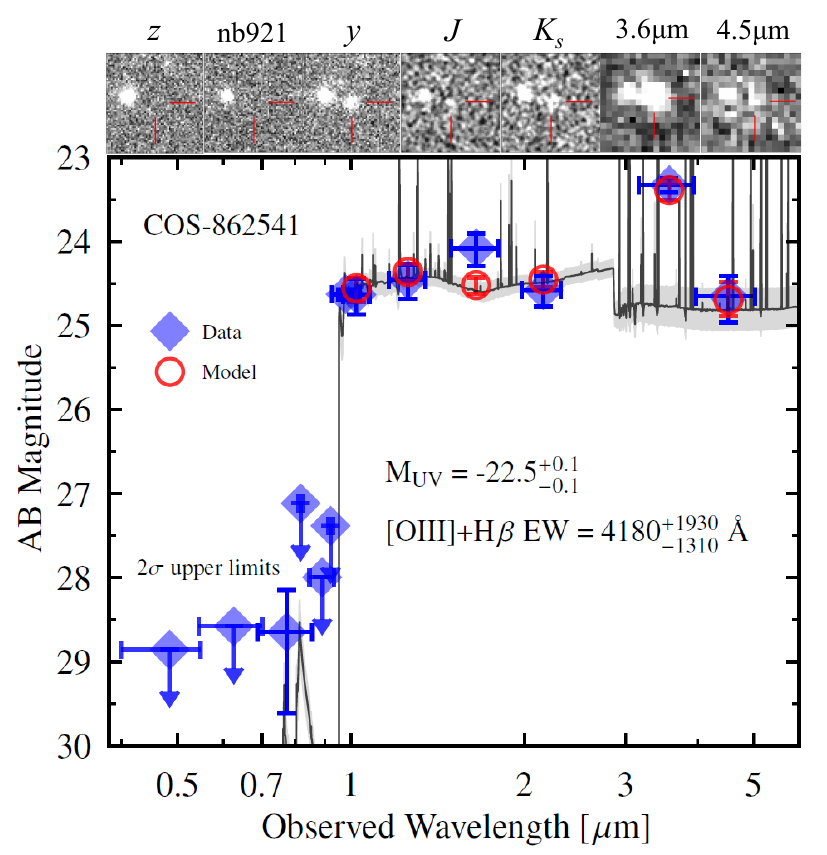}
\caption{\textbf{Top:} HSC \textit{z},nb921,\textit{y}, VIRCam \textit{J},\textit{K}$_s$, and IRAC 3.6$\mu$m and 4.5$\mu$m postage stamp images (10\arcsec{} $\times$ 10\arcsec{}) of COS-862541. The source is clearly seen to be a sharp nb921 dropout possessing a strong 3.6$\mu$m excess. \textbf{Bottom:} Spectral energy distribution of COS-862541 with the BEAGLE SED model fit adopting a systemic redshift of \zsys{} = 6.837 (see text) overlaid. The format of this panel is very similar to those in Fig. \ref{fig:variousSEDs}.}
\label{fig:BEAGLE_fit_COS-862541}
\end{figure}

\begin{table}
\centering
\caption{Measured aperture photometry for COS-862541 in Subaru/HSC \textit{g},\textit{r},\textit{i},nb816,\textit{z},nb921,\textit{y}, VISTA/VIRCam \textit{Y},\textit{J},\textit{H},\textit{K}$_s$, and IRAC 3.6 and 4.5$\mu$m bands. We report 2$\sigma$ limiting magnitudes for bands where the source is undetected.}
\begin{tabular}{P{0.6cm}P{1.2cm}P{1.1cm}P{0.6cm}P{1.3cm}P{1.1cm}}
\multicolumn{6}{c}{COS-862541 Photometry} \\
\hline
Band & Flux (nJy) & AB Mag & Band & Flux (nJy) & AB Mag\\[2pt]
\hline
\textit{g} & 3 $\pm$ 5 & $>$28.86 & \textit{Y} & 511 $\pm$ 99 & 24.63$^{+0.24}_{-0.19}$ \\[2pt]
\textit{r} & $-$7 $\pm$ 7 & $>$28.58 & \textit{J} & 601 $\pm$ 112 & 24.45$^{+0.22}_{-0.18}$\\[2pt]
\textit{i} & 13 $\pm$ 7 & 28.65$^{+0.94}_{-0.50}$ & \textit{H} & 846 $\pm$ 151 & 24.08$^{+0.21}_{-0.18}$\\[2pt]
nb816 & 8 $\pm$ 26 & $>$27.12 & \textit{K}$_s$ & 536 $\pm$ 89 & 24.58$^{+0.20}_{-0.16}$\\[2pt]
\textit{z} & 4 $\pm$ 12 & $>$27.99 & 3.6$\mu$m & 1697 $\pm$ 133 & 23.33$^{+0.09}_{-0.08}$\\[2pt]
nb921 & $-$6 $\pm$ 20 & $>$27.38 & 4.5$\mu$m & 500 $\pm$ 124 & 24.65$^{+0.31}_{-0.24}$\\[2pt]
\textit{y} & 513 $\pm$ 27 & 24.62$^{+0.06}_{-0.06}$ & - & -  & - \\[2pt]
\hline
\end{tabular}
\label{tab:COS-862541_photometry}
\end{table}

\subsection{Rest-UV Spectroscopy} \label{sec:spec_confirmation}

We have initiated a survey to search for \Lya{} in this sample. Our first set of observations included only one candidate, COS-862541. We prioritized COS-862541 due to its bright UV luminosity (\textit{J} = 24.45) and very blue [3.6]$-$[4.5] colour of $-$1.33$^{\scaleto{+0.26}{4.5pt}}_{\scaleto{-0.32}{4.5pt}}$, suggesting an extreme rest-frame [OIII]+H$\beta$ EW $\sim$ 2250 \AA{} (see Table \ref{tab:galaxy_properties}). We show postage stamp images of COS-862541 in the top panel of Fig. \ref{fig:BEAGLE_fit_COS-862541} where it is clearly seen to be a strong nb921 dropout possessing a very strong [3.6] excess. We list the full optical through mid-infrared photometry for COS-862541 in Table \ref{tab:COS-862541_photometry}.

We observed COS-862541 using the MMT/Binospec spectograph \citep{Fabricant2019} on the night of 13 December 2018 for 2.0 hours (eight 900 second exposures). Conditions were clear and the average seeing was 1.1\arcsec{}. We used the \textsc{BinoMask} software to design this mask using a slit width of 1.0\arcsec{}, resulting in a resolving power of $R \approx 4360$. Spectra were taken using the 600 l/mm grating adopting a central wavelength of 0.85$\mu$m yielding a wavelength coverage of 7250--9775 \AA{} for COS-862541, corresponding to 4.96 $<$ \zLya{} $<$ 7.04. Off the primary FoVs, the mask included five stars for guiding and monitoring seeing throughout the observations. We placed slits on three additional stars within the primary FoVs for absolute flux calibration. The resulting Binospec spectra were reduced using the publicly available data reduction pipeline \citep{Kansky2019}. Because Binospec is a multi-object spectrograph, we were able to simultaneously observe approximately one hundred z$\sim$5--6 candidates. The results for these sources will be described in a future paper after more observations have been conducted.

We extracted a one-dimensional spectrum in a rectangular aperture $\pm$3$\sigma$ wide from the 1.1\arcsec{} seeing (corresponding to $\pm$6 spatial pixels). We then performed absolute flux calibration by determining the average scaling factor that matches the 1D spectra of the three stars placed on the mask with their mean PSF $z$-band magnitudes from the Pan-STARRS survey \citep{Chambers2016_PanSTARRS}. Given the relatively narrow wavelength range covered by these observations ($\sim$0.7--1$\mu$m), we assume that this factor does not evolve with wavelength. To estimate slit loss, we assume that the surface brightness of COS-862541 follows a S\'ersic profile with $n = 1.5$ \citep[e.g.][]{Oesch2010,Ono2013} and a half-light radius of $r_e = 2.30$ kpc following the morphological study of similarly luminous z$\sim$7 galaxies in \citet{Bowler2017}. This S\'ersic profile is then convolved with a 1.1\arcsec{} FWHM 2D Gaussian and the fraction of flux within the 2D aperture is compared to that for a point source. The relative slit loss correction factor is calculated to be $\approx$1.25. To determine the flux errors, we add 10,000 realizations of the noise to each pixel within the extracted 2D aperture, compute the measured flux for each realization, and calculate the standard deviation from the resulting distribution.

\begin{figure}
\includegraphics{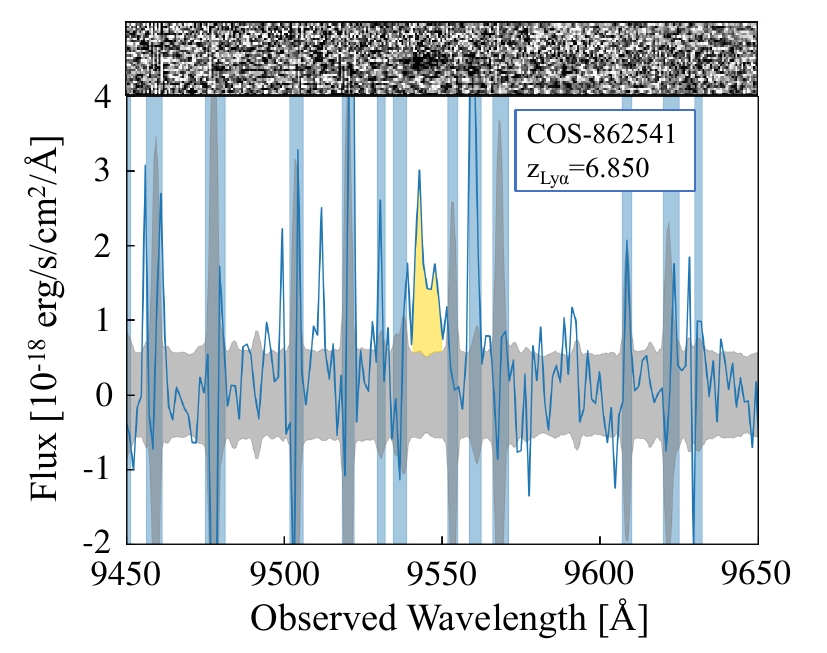}
\caption{MMT/Binospec spectra of COS-862541 wherein an asymmetric emission feature is revealed at the spatial position of the source. We interpret this feature as \Lya{} at z=6.850. The top panel shows the 2D signal-to-noise ratio map where black is positive while the bottom panel shows the flux calibrated 1D extraction with the 1$\sigma$ noise level in gray. OH skyline locations are shaded over in blue.}
\label{fig:COS-862541_Binospec_Spectrum}
\end{figure}

The MMT/Binospec observations revealed a highly significant asymmetric feature at 9543 \AA{} which we identify as \Lya{} (see Fig. \ref{fig:COS-862541_Binospec_Spectrum}). We measure a line flux of $\left(20.0 \pm 1.9\right) \times 10^{-18}$ erg/s/cm$^2$ and calculate the redshift to be \zLya{} = 6.850 using the location of peak line flux. Because the continuum is undetected in the spectrum, we calculate the rest-frame EW using a continuum flux set by the VIRCam \textit{Y}-band photometry. This band is that closest to a rest-frame wavelength of 1215.67 \AA{} while not being contaminated by \Lya{}. The resulting rest-frame \Lya{} EW is 15 $\pm$ 3 \AA{}, comparable to that of previously discovered z$\gtrsim$7 \Lya{} emitters with strong IRAC colours \citep{Ono2012,Finkelstein2013,Oesch2015,Zitrin2015,RobertsBorsani2016,Stark2017,Jung2019}.

Now informed with the redshift of COS-862541, we can reassess the inferred [OIII]+H$\beta$ EW. At z=6.850, [OIII]$\lambda$5007 lies at an observed wavelength with 22.6\% maximum transmission through the 3.6$\mu$m filter meaning that the [OIII]$+$H$\beta$ EW inferred for this source in \S\ref{sec:galaxy_properties} is likely too small (the photometric redshift was inferred to be z=6.72 where [OIII]$\lambda$5007 is near maximum transmission through [3.6]). However, because \Lya{} is often observed to be redshifted relative to systemic in luminous high-redshift galaxies \citep[e.g.][]{Willott2015,Stark2017,Hutchison2019}, it is likely that the true redshift of COS-668525 is overestimated and therefore the transmission of [OIII]$\lambda$5007 underestimated. To account for this uncertainty, we consider two extreme cases. First, we consider $\Delta$v$_{\mathrm{Ly\alpha}}$ = 0 km s$^{-1}$ resulting in a systemic redshift, \zsys{}, equal to \zLya{}. Second, we consider a \Lya{} velocity offset of $\Delta$v$_{\mathrm{Ly\alpha}}$ = 500 km s$^{-1}$, the maximum yet observed for z$\sim$7 galaxies \citep{Willott2015}. This results in \zsys{} = 6.837 and boosts the transmission of [OIII]$\lambda$5007 through the 3.6$\mu$m filter to 35.6\%. 

We re-fit the VIRCam and IRAC photometry of COS-862541 with BEAGLE\footnote{We do not use any of the HSC photometry in this re-fitting process as all HSC bands either bluewards of or contaminated by \Lya{} at the redshift of COS-862541 and it is not possible to modulate the effective \Lya{} transmission fraction in BEAGLE.} adopting \zsys{} = 6.837 and \zsys{} = 6.850. We infer an [OIII]$+$H$\beta$ rest-frame EW of 4180$^{\scaleto{+1550}{4.5pt}}_{\scaleto{-1310}{4.5pt}}$ \AA{} and 4880$^{\scaleto{+1630}{4.5pt}}_{\scaleto{-1640}{4.5pt}}$ \AA{} for \zsys{} = 6.837 and \zsys{} = 6.850, respectively. Both redshifts yield \Muv{} = $-$22.5$^{\scaleto{+0.1}{4.5pt}}_{\scaleto{-0.1}{4.5pt}}$ from the posterior distribution output by BEAGLE. Because two of three \Muv{} $<$~$-$22.4 galaxies at z$>$6 with \Lya{} velocity offset measurements have $\Delta$v$_{\mathrm{Ly\alpha}}$ $>$ 400 km s$^{-1}$ \citep{Willott2015,Stark2017}, we adopt the inferred properties with \zsys{} = 6.837 (v$_{\mathrm{Ly\alpha}}$ = 500 km s$^{-1}$) values as fiducial but fold the v$_{\mathrm{Ly\alpha}}$ uncertainty into the reported uncertainty on inferred parameters. Doing so yields an inferred \Muv{} = $-$22.5$^{\scaleto{+0.1}{4.5pt}}_{\scaleto{-0.1}{4.5pt}}$ and [OIII]$+$H$\beta$ rest-frame EW of 4180$^{\scaleto{+1930}{4.5pt}}_{\scaleto{-1310}{4.5pt}}$ \AA{}. 

Such extreme emission properties can be reproduced (within the context of BEAGLE) by galaxies possessing moderate metallicity (0.23$^{\scaleto{+0.19}{4.5pt}}_{\scaleto{-0.09}{4.5pt}}$ Z$_{\odot}$) and extremely large sSFR (178$^{\scaleto{+238}{4.5pt}}_{\scaleto{-136}{4.5pt}}$ Gyr$^{-1}$) where the emergent starlight is dominated by a very young ($\sim$3 Myr) stellar population.
While this is the largest [OIII]$+$H$\beta$ EW inferred from a spectroscopically confirmed z$\gtrsim$7 galaxy, it is still consistent with that measured in extreme emission line galaxies at z$\lesssim$2 \citep{Chevallard2018_z0,Izotov2018b,Tang2019}. 
Because [OIII] EW has been shown to strongly correlate with Lyman continuum production efficiency (at least at z$\lesssim$3; \citealt{Chevallard2018_z0}; \citealt{Tang2019}; \citealt{Emami2020}; \citealt{Nakajima2020}), this suggests that COS-862541 is one of the largest producers of ionizing photons in the reionization era.
With an absolute UV magnitude of \Muv{} = $-$22.5$^{\scaleto{+0.1}{4.5pt}}_{\scaleto{-0.1}{4.5pt}}$, COS-862541 is also one of the most luminous galaxies exhibiting visible \Lya{} at z$>$6.8. 
Recent work investigating the spectral properties of very bright z$\gtrsim$7 \Lya{} emitters with strong ($\gtrsim$800 \AA{} EW) [OIII]$+$H$\beta$ emission has often uncovered prominent rest-UV metal line emission, providing insight into the nature of these sources \citep{Stark2017,Laporte2017,Mainali2018}. 
COS-862541 is therefore an excellent candidate for spectroscopic follow-up to help clarify the nature of very bright \Lya{} emitters in the reionization era. 
We estimate a stellar mass of \Mstar{} = $\left(1.1^{\scaleto{+2.2}{4.5pt}}_{\scaleto{-0.6}{4.5pt}}\right) \times 10^9$ \Msol{}, suggesting that COS-862541 is caught in a phase with relatively low \Mstar{}/L$_{\mathrm{UV}}$ ratio compared to trends from empirical \Muv{}-\Mstar{} relations at z$\sim$7 \citep[e.g.][]{Duncan2014,Song2016,Behroozi2019}. 

\section{Analysis} \label{sec:analysis}

In this section, we infer the [OIII]$+$H$\beta$ EW distribution at z$\simeq$7 using a ground-based sub-sample of 20 galaxies (COSMOS+XMM1; \S\ref{sec:COSMOSXMM1_Selection}) and an \HST{}-based sub-sample of 22 galaxies (GOODS; \S\ref{sec:GOODS_Selection}). The COSMOS+XMM1 galaxies were identified using a novel combination of both narrow and broad-band filters to precisely target the redshift range z$\simeq$6.63--6.83 where very blue [3.6]$-$[4.5] IRAC colours unambiguously indicate strong [OIII]$+$H$\beta$ emission. For this reason, we adopt the results from the COSMOS+XMM1 sub-sample as fiducial and use the generally fainter GOODS sub-sample to investigate whether there is any evidence that the [OIII]$+$H$\beta$ EW distribution varies strongly with UV luminosity. We characterize the [OIII]$+$H$\beta$ EW distribution with two parameters, the median EW, \muEW{}, and the standard deviation, \sigmaEW{}. In doing so, we assume that the z$\simeq$7 [OIII]$+$H$\beta$ EW distribution is log-normal, motivated by the distribution shape for H$\alpha$ at lower redshifts \citep{Lee2007,Lee2012,Ly2011}. To infer \muEW{} and \sigmaEW{}, we adopt two different methods each explained in turn below. By taking two different approaches, we are able to assess the systematic uncertainties on our inferred distribution. We report the inferred [OIII]$+$H$\beta$ EW distributions after detailing both methods.

\textit{Method 1}: For our first approach, we use the [OIII]$+$H$\beta$ EWs inferred by the BEAGLE SED model fits described in \S\ref{sec:galaxy_properties}. As detailed therein, the typical inferred [OIII]$+$H$\beta$ EW uncertainty for a given source is 0.33 dex though sources with the highest inferred EWs tend to have the smallest uncertainty ($\lesssim$0.2 dex for EW $>$1200 \AA{} sources). Fig. \ref{fig:EWposteriors} shows the BEAGLE posterior [OIII]$+$H$\beta$ EW distributions in a representative set of our sample.

To infer the [OIII]$+$H$\beta$ EW distribution, we generate a grid spanning \logten{}(\muEW{}/\AA{}) = 1.0--3.5 and \sigmaEW{} = 0.01--1.0 dex with a spacing of 0.01 dex for both parameters. We then compute the probability for a given set of parameters, $P(\muEW{},\sigmaEW{})$, following a simplified version of the approach in Boyett et al. (2020, in prep):
\begin{equation} \label{eq:prob}
P \left(\muEW{},\sigmaEW{}\right) \propto \prod_i \int P_i\left(\mathrm{EW}\right) P\left(EW\ |\ \muEW{},\sigmaEW{}\right)\  d\mathrm{EW}.
\end{equation}
Here, $P_i(\mathrm{EW})$ is the probability distribution function (PDF) of the [OIII]$+$H$\beta$ EW for source $i$ as inferred from BEAGLE, and $P(EW | \muEW{},\sigmaEW{})$ is the log-normal [OIII]$+$H$\beta$ EW distribution with parameters \muEW{} and \sigmaEW{}. The expression in the integral is appropriately small when the assumed log-normal EW distribution is inconsistent with the BEAGLE posterior PDF and large when the two distributions are similar. Furthermore, sources with very broad [OIII]$+$H$\beta$ EW PDFs from BEAGLE will appropriately have relatively poor constraining power while those with precise PDFs will largely determine $P(\muEW{},\sigmaEW{})$. We marginalize over the full grid to determine $P(\muEW{})$ and $P(\sigmaEW{})$ and report the median and 68\% confidence interval values on each parameter. Due to our spectroscopic \Lya{} detection for COS-862541, we use the posterior of [OIII]$+$H$\beta$ EWs when fixing z=6.837 ($\Delta$v$_{\mathrm{Ly\alpha}}$ = 500 km s$^{-1}$; see \S\ref{sec:spec_confirmation}) during the BEAGLE fit for this source. Our results change negligibly if we instead adopt z=6.850 ($\Delta$v$_{\mathrm{Ly\alpha}}$ = 0 km s$^{-1}$).

\textit{Method 2}: In our second method, we determine $P_i(\mathrm{EW})$ directly from the measured IRAC colour distribution and expected redshift distribution of each source using a forward modeling approach. The measured IRAC colour distribution, $P_i([3.6]-[4.5])$, is generated by adding realizations of the measured noise to the IRAC fluxes for source $i$ and computing the resulting colour in each realization. We also account for the dependence of redshift on the mapping between [3.6]$-$[4.5] colour and [OIII]$+$H$\beta$ EW by sampling redshifts from the selection completeness simulations described in \S\ref{sec:sample_selection}. 
Given the $\sim$10\% fraction of strong \Lya{} emitters (\Lya{} EW $>$ 25 \AA{}) at z$\sim$7 \citep[e.g.][]{Pentericci2018}, we build the expected redshift distribution for each source, $P_i(\mathrm{z})$, by sampling redshifts 90\% of the time from the completeness simulations assuming \Lya{} EW = 0 \AA{} and 10\% of the time assuming \Lya{} EW = 25 \AA{}. Finally, we account for the impact of dust on the rest-optical continuum and, therefore, the resulting [3.6]$-$[4.5] colour at fixed [OIII]$+$H$\beta$ EW. 

While the majority of the sources in our sample have little inferred dust attenuation ($\tau_{\mathrm{V}} < 0.05$), there are some sources with red rest-UV slopes ($\beta > -1.5$) suggesting significant attenuation ($\tau_{\mathrm{V}} \gtrsim 0.1$). We therefore adjust our modeled rest-optical continuum slope based on the inferred dust reddening. To do so, we adopt an SMC attenuation law \citep{Pei1992}, motivated by recent IRX-$\beta$ measurements at z$\sim$2--3 \citep{Bouwens2016_ALMA,Reddy2018_dust}, and assume an intrinsically flat continuum between the rest-optical wavelengths covered by [3.6] and [4.5], consistent with the typical un-attenuated rest-optical slope in our sample as inferred by BEAGLE. $A_V$ values are pulled from the posterior PDF output by BEAGLE for each source when excluding IRAC photometry from the fit. By excluding IRAC photometry, we remove any bias on the inferred dust reddening imposed by strong IRAC colours. 

$P_i(\mathrm{EW})$ is therefore calculated by sampling IRAC colours from $P_i([3.6]-[4.5])$, redshifts from $P_i(\mathrm{z})$, and rest-optical slopes using $P_i(A_V)$. When mapping a [3.6]$-$[4.5] colour to a (redshift and dust dependent) [OIII]$+$H$\beta$ EW, we also consider the minor contribution from fainter nebular emission lines (H$\alpha$, H$\gamma$, H$\delta$, [OII]$\lambda$3727,3729, [NeIII]$\lambda$3869, [NII]$\lambda$6548,6583, and [SII]$\lambda$6716,6730) adopting the empirical and theoretical line ratios described in Appendix \ref{appendix:line_ratios}. $P(\muEW{},\sigmaEW{})$ is then calculated using Eq. \ref{eq:prob}, adopting the same grid on the log-normal EW parameters as for Method 1.

\begin{table}
\centering
\caption{Inferred values for the median EW, \muEW{}, and standard deviation, \sigmaEW{}, characterizing the z$\simeq$7 [OIII]$+$H$\beta$ EW distribution. We report values inferred for the COSMOS+XMM1 and GOODS sub-samples separately, as well as when we combine both sub-samples (All). We also report values inferred from each approach (Method 1 and Method 2) detailed in \S\ref{sec:analysis} as well as the values obtained by convolving the probability distribution functions from each approach (Method 1+2). Reported values and uncertainties are the median and 68\% confidence intervals, respectively, from the posterior distribution functions.}
\begin{tabular}{P{2cm}P{1.3cm}P{1.3cm}P{1.5cm}}
\multicolumn{4}{c}{\logten{}(\muEW{}/\AA{})} \\
\hline
Sample & Method 1 & Method 2 & Method 1$+$2\Tstrut{}\\[2pt]
\hline
COSMOS+XMM1 & 2.85$^{+0.08}_{-0.08}$ & 2.92$^{+0.10}_{-0.11}$ & 2.88$^{+0.06}_{-0.07}$ \Tstrut{}\\[2pt]
GOODS & 2.85$^{+0.06}_{-0.07}$ & 2.92$^{+0.09}_{-0.10}$ & 2.87$^{+0.05}_{-0.05}$\\[2pt]
All & 2.85$^{+0.05}_{-0.05}$ & 2.92$^{+0.07}_{-0.08}$ & 2.87$^{+0.04}_{-0.04}$\\[2pt]
\hline
\end{tabular}
\begin{tabular}{P{2cm}P{1.3cm}P{1.3cm}P{1.5cm}}
 & & & \\
\end{tabular}
\begin{tabular}{P{2cm}P{1.3cm}P{1.3cm}P{1.5cm}}
\multicolumn{4}{c}{\sigmaEW{} [dex]} \\
\hline
Sample & Method 1 & Method 2 & Method 1$+$2\Tstrut{}\\[2pt]
\hline
COSMOS+XMM1 & 0.24$^{+0.09}_{-0.07}$ & 0.30$^{+0.11}_{-0.08}$ & 0.26$^{+0.06}_{-0.05}$ \Tstrut{}\\[2pt]
GOODS & 0.14$^{+0.08}_{-0.08}$ & 0.22$^{+0.10}_{-0.10}$ & 0.16$^{+0.07}_{-0.05}$\\[2pt]
All & 0.19$^{+0.05}_{-0.05}$ & 0.26$^{+0.06}_{-0.06}$ & 0.21$^{+0.05}_{-0.03}$\\[2pt]
\hline
\end{tabular}
\label{tab:inferredEWparameters}
\end{table}

\begin{figure}
\includegraphics{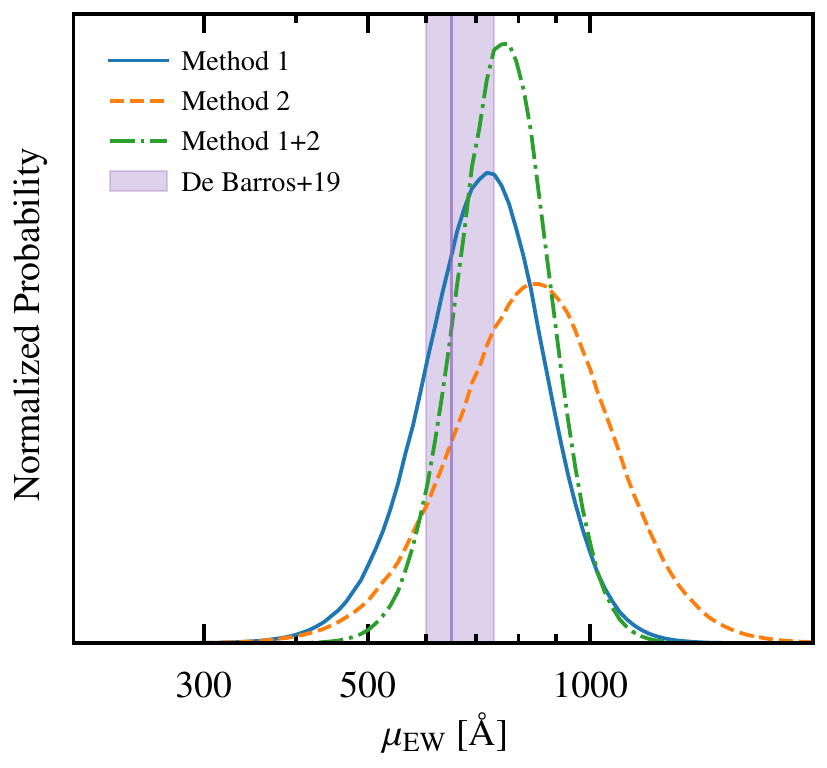}
\includegraphics{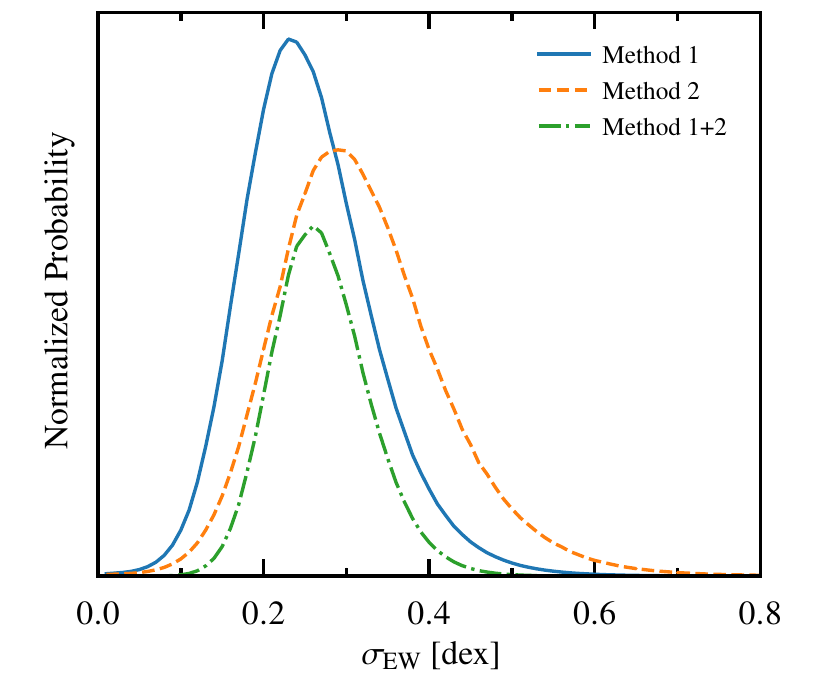}
\caption{Inferred normalized PDFs of the median EW, \muEW{} (top), and standard deviation, \sigmaEW{} (bottom), of the [OIII]$+$H$\beta$ EW distribution for our COSMOS+XMM1 sub-sample selected to lie at z$\simeq$6.63--6.83. The solid blue curves show the PDFs obtained using the posterior [OIII]$+$H$\beta$ EWs output by the SED fitting code BEAGLE (Method 1) while the dashed orange curves show the PDFs obtained by using the observed IRAC [3.6]$-$[4.5] colours of each source in a forward-modeling approach (Method 2). The dot-dashed green curves show the PDFs calculated by convolving the PDFs from the two methods (Method 1+2). In the top panel, we also show the median [OIII]$+$H$\beta$ EW at z$\simeq$8 reported by \citet{deBarros2019} which is consistent with our inferred median EW at z$\simeq$6.63--6.83.}
\label{fig:AnalysisFigure}
\end{figure}

\begin{figure}
\includegraphics{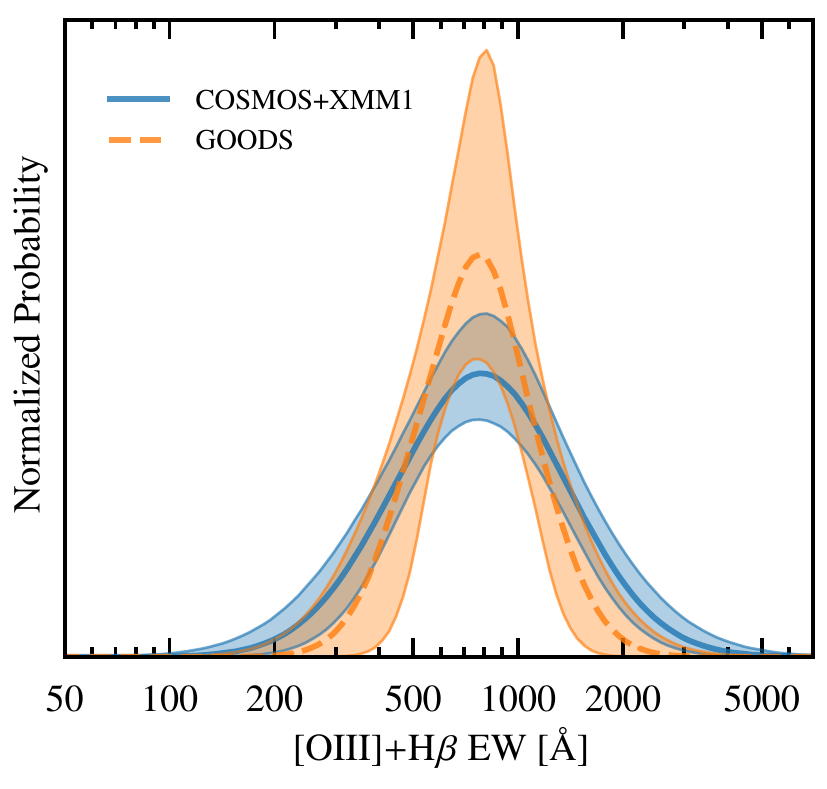}
\caption{Inferred z$\simeq$7 [OIII]$+$H$\beta$ EW distributions using the COSMOS+XMM1 (blue) and GOODS (orange) sub-samples. Here, we adopt \muEW{} and \sigmaEW{} values from the combined posteriors of each approach (Method 1+2). The inferred distributions for both sub-samples are consistent within one another within uncertainties (shaded regions).}
\label{fig:finalEWdistn}
\end{figure}

We report the inferred \muEW{} and \sigmaEW{} values from both methods in Table \ref{tab:inferredEWparameters}. For the COSMOS+XMM1 sub-sample, we infer a median EW of \logten{}(\muEW{}/\AA{}) = 2.85$^{\scaleto{+0.08}{4.5pt}}_{\scaleto{-0.08}{4.5pt}}$ and 2.92$^{\scaleto{+0.10}{4.5pt}}_{\scaleto{-0.11}{4.5pt}}$ and a standard deviation of 0.24$^{\scaleto{+0.09}{4.5pt}}_{\scaleto{-0.07}{4.5pt}}$ and 0.30$^{\scaleto{+0.11}{4.5pt}}_{\scaleto{-0.08}{4.5pt}}$ dex with Method 1 and Method 2, respectively. Therefore, we find that the two methods deliver consistent values within statistical uncertainties. To enable a more direct comparison, we plot the PDFs on \muEW{} and \sigmaEW{} for both methods in Fig. \ref{fig:AnalysisFigure} which illustrates that the PDFs from both methods overlap significantly indicating consistency. 

To infer \muEW{} and \sigmaEW{} values, we choose to convolve the PDFs from Method 1 and Method 2.
In doing so, we directly account for systematic uncertainties and obtain an EW distribution most consistent with both approaches which each have their own advantages.
The primary advantage of Method 1 is that BEAGLE allows for more flexibility in the possible SED shapes and hence the inferred PDFs of [OIII]$+$H$\beta$ EWs.
Method 2, on the other hand, allows us to consider the impact of \Lya{} on the redshift distribution of our sources.

As shown in Fig. \ref{fig:AnalysisFigure}, the convolved PDFs from both approaches (Method 1+2) are highly consistent with that from each individual approach and result in slightly higher precision on the log-normal EW parameters. For the COSMOS+XMM1 sub-sample (from which we take fiducial values), this combined PDF yields \muEW{} = 759$^{\scaleto{+112}{4.5pt}}_{\scaleto{-113}{4.5pt}}$ \AA{} and \sigmaEW{} = 0.26$^{\scaleto{+0.06}{4.5pt}}_{\scaleto{-0.05}{4.5pt}}$ dex.
This distribution is shown in Fig. \ref{fig:finalEWdistn}.
At z$\simeq$8, \citet{Labbe2013} infer a typical [OIII]$+$H$\beta$ EW = 670$^{\scaleto{+260}{4.5pt}}_{\scaleto{-170}{4.5pt}}$ \AA{} using the average [3.6]$-$[4.5] colour from 20 \HST{}-selected galaxies. More recently, \citet{deBarros2019} fit the SEDs of 76 \HST{}-selected galaxies at z$\simeq$8 to find a median EW of 649$^{\scaleto{+49}{4.5pt}}_{\scaleto{-52}{4.5pt}}$ \AA{}. Both values are consistent with our median [OIII]$+$H$\beta$ EW at z$\simeq$6.63--6.83 within uncertainties, suggesting that the typical [OIII]$+$H$\beta$ EW does not evolve strongly in the 170 Myr between z$\sim$6.7--8. 

However, we still find that the [OIII]$+$H$\beta$ EW distribution has evolved substantially from z$\lesssim$2. 
As stated in \citet{deBarros2019}, only 0.23\% of galaxies in the Sloan Digital Sky Survey (SDSS; \citealt{Abolfathi2018}) have [OIII]$+$H$\beta$ EWs $\geq$ 300 \AA{}. 
Results from our COSMOS+XMM1 sub-sample suggest that approximately 90\% of z$\simeq$7 galaxies possess such prominent rest-optical emission. 
Even massive (\Mstar{}$\sim$10$^{10}$ \Msol{}) star-forming galaxies at z$\simeq$2.3 only have a typical [OIII]$+$H$\beta$ EW = 110 \AA{} \citep{Reddy2018_opticalLines} which is $>$5$\times$ less than the typical EW we infer here. 
Moreover, we find that a substantial fraction\footnote{This value was calculated by sampling from the Method 1+2 $P(\muEW{},\sigmaEW{})$ grid of the COSMOS+XMM1 sub-sample. The inferred fraction becomes 18$^{\scaleto{+6}{4.5pt}}_{\scaleto{-5}{4.5pt}}$\% if we include the GOODS sub-sample.} (23$\pm$7\%) of z$\simeq$7 galaxies possess [OIII]$+$H$\beta$ EWs in excess of 1200 \AA{}. 
Such extreme emission is very rare in the massive z$\simeq$2 galaxy population \citep{Reddy2018_opticalLines}.

We find no evidence of strong evolution in the z$\simeq$7 [OIII]$+$H$\beta$ EW distribution with UV luminosity. As can be seen in Table \ref{tab:inferredEWparameters} and Fig. \ref{fig:finalEWdistn}, both the \muEW{} and \sigmaEW{} values inferred from the bright (\Muv{} $\lesssim$ $-$21) COSMOS+XMM1 sub-sample are consistent with those inferred from the fainter (\Muv{} $\lesssim$ $-$20) GOODS sub-sample within uncertainties. 
Studies at z$\sim$1--3 have found that [OIII]$+$H$\beta$ EWs increase steadily with decreasing stellar mass, specifically evolving as EW $\propto$ \Mstar{}$^{\alpha}$ where $\alpha \sim -0.35$ \citep{Khostovan2016}. 
Using BEAGLE, we find that the typical inferred stellar mass of the COSMOS+XMM1 and GOODS sub-samples are \logMstar{} = 8.80 and 8.65, respectively, suggesting that the typical EWs of these sub-samples should differ by $\sim$0.05 dex.
We infer a difference of $-$0.01$\pm$0.08 dex so, within uncertainties, our results are consistent with a similar \Mstar{}$-$EW relation at z$\simeq$7.
None the less, the large scatter in the \Muv{}-\Mstar{} relation at z$\sim$6$-$7 ($\sim$0.3--0.4 dex; \citealt{Duncan2014,Salmon2015,Song2016}) suggests an even weaker evolution of EW with UV luminosity. 
Upcoming \JWST{} surveys will soon clarify this relation by delivering [OIII]$+$H$\beta$ EW measurements for fainter high-redshift galaxies and by improving EW constraints for galaxies with UV luminosities studied here.

The median [OIII]$+$H$\beta$ EW we derive (759 \AA{}) is consistent with that reported in a smaller sample of lensed z$\simeq$6.8 galaxies ($>$637 \AA{}) from \citet{Smit2014}. 
We can also compare to the results of \citet{Smit2015} after accounting for differences in IRAC selection criteria. 
Therein, they pre-select z$>$6.5 CANDELS galaxies with blue IRAC colors (and hence large [OIII]$+$H$\beta$ EWs) by adopting P([3.6]$-$[4.5] $<$ $-$0.5) $>$ 0.84 and find that the median inferred [OIII]$+$H$\beta$ EW is 1375 \AA{}\footnote{Here we are reporting the median EW from \citet{Smit2015} before they account for the bias due to noise in measured IRAC fluxes. We do so to better compare with our results. The median EW found by \citet{Smit2015} after correcting for this noise bias was 1085 \AA{}.} among these sources. 
If we apply the same blue IRAC color selection criteria to our COSMOS+XMM1 sub-sample (ignoring the $f_{J}/e_{[4.5]}>2$ cut) and use the Method 1 approach, we obtain a comparable median EW of 1622 \AA{}. 
Finally, we note that our EW distribution is consistent with the range of [OIII]$+$H$\beta$ EWs inferred from the stacking analysis of \citet{Castellano2017}.

\section{Discussion} \label{sec:discussion}

Our understanding of early galaxies has advanced considerably over the past decade.
Deep imaging surveys conducted by \HST{} and \Spitzer{} have revealed that star forming systems in the reionization era are different from those at z$\sim$2, with lower masses \citep{Duncan2014,Song2016}, larger sSFRs \citep{Sanders2018,Strom2018,Duncan2014,Song2016}, and smaller sizes \citep{Shibuya2015}. 
In this paper, we demonstrate that the [OIII]+H$\beta$ EW  distribution evolves substantially as well, shifting toward larger values (see also \citealt{Labbe2013,Smit2014,Smit2015,deBarros2019}).  
We find that typical z$\simeq$7 systems have [OIII]+H$\beta$ EW = 759 \AA{}, implying both large sSFR (4.4 Gyr$^{-1}$) and moderately low metallicities (0.16 Z$_\odot$).
We also find evidence for the emergence of a yet more extreme population ([OIII]+H$\beta$ EW $>$ 1200~\AA) that is rarely seen in similarly selected samples at lower redshift. 
These are compact galaxies with very high star formation for their mass (sSFR = 36 Gyr$^{-1}$), implying extremely large star formation rate surface densities. 
Given the very small mass-doubling times, these galaxies are likely not in this large sSFR phase for very long, as might be expected for galaxies undergoing a burst of star formation. 
None the less, given that $\approx$20\% of the z$\simeq$7 population possesses such intense line emission (see \S\ref{sec:analysis}), galaxies must cycle through this phase of intense star formation regularly in the reionization era.

In this section, we explore these extreme sSFR galaxies in more detail, considering, in turn, their efficiency in forming star clusters and ionizing the IGM.  
Attention has recently focused on their rest-UV properties, with targeted spectroscopic programs revealing high ionization metal lines (i.e. He II, CIV, CIII]) that are rarely seen in galaxies that are typical at z$\simeq$2--3. 
The prominence of these lines indicates hard ionizing spectra powered by dominant populations of very young and hot massive stars (e.g. \citealt{Stark2015_CIV,Mainali2017,Vanzella2019}).  
The magnification from gravitational lensing provides a sharper view, resolving the stellar populations of extreme sSFR galaxies at z$\sim$2$-$6 into compact ($<$20 pc) and dense clusters \citep{RiveraThorsen2017,Vanzella2017,Vanzella2019,Vanzella2020}. 
The implied star formation rate surface densities in these clusters are extremely large, consistent with formation scenarios for globular clusters \citep{Vanzella2019}.  
Among nearby galaxies, the efficiency of forming bound clusters is known to scale with star formation rate surface density (e.g. \citealt{Bastian2008,Goddard2010,Adamo2011}), reaching very large values ($\simeq$50\%) in blue compact dwarf starbursts, a population that is very similar to the extreme optical line emitters ([OIII]+H$\beta$ EW$>$1200~\AA) described above. 
If this trend holds at high redshift, we would expect galaxies to be very effective factories of star cluster formation as they pass through this very high sSFR phase. 
The stellar mass inferred to have formed in the past 10 Myr for the most extreme sSFR galaxies in our sample (COS-862541 and COS-1099982) is $\sim$3$\times$10$^8$ \Msol{} in the context of BEAGLE, meaning that each of these sources very well could have recently formed multiple super star clusters.
The rapid rise in the prevalence of the most extreme optical line emitters at z$>$6 may thus go hand in hand with a rise in the formation efficiency of dense star clusters.  

The changing demographics of galaxy populations also has implications for the contribution of galaxies to reionization. 
The first detections of intense UV metal nebular emission in z$\simeq$7--8 galaxies with extremely large [OIII]+H$\beta$ EW led to suggestions that this population is likely to be extremely efficient in producing hydrogen ionizing photons (e.g. \citealt{Stark2017}). 
This has since been shown more systematically using larger samples at lower redshifts \citep{Chevallard2018_z0,Tang2019,Emami2020,Nakajima2020}. 
These studies clearly demonstrate that the production efficiency of ionizing radiation increases with [OIII]+H$\beta$ EW, reaching very large values in the 
most extreme line emitters. 
This trend reflects a shift toward hotter and younger stellar populations at larger [OIII]+H$\beta$ EW.
With the overall galaxy population evolving toward larger [OIII]+H$\beta$ EWs at earlier times, we expect ionizing photon production to become more efficient as we enter the reionization era.  
The most prodigious ionizing agents will be those systems caught during an extreme 
line emitting phase. 
Assuming the trend between $\xi_{\rm{ion}}$ and [OIII] EW derived in \citet{Tang2019} holds at z$\simeq$7, those galaxies with [OIII]+H$\beta$ EW=1200--3000~\AA\ will produce 2.6--5.2$\times$ the number of ionizing photons (relative to their UV continuum luminosity at 1500~\AA) as typical massive (\Mstar{}=10$^{10}$ \Msol{}) galaxies at z$\simeq$2 \citep{Shivaei2018}.

But regardless of how efficiently a galaxy can produce ionizing photons, it will only contribute to reionization if it leaks its Lyman-continuum (LyC) radiation into the IGM. 
Over the past five years, the first statistical samples of LyC emitters have begun to emerge at lower redshift, providing insight into the properties that facilitate ionizing photon escape \citep[e.g.][]{Chisholm2017,Steidel2018,Izotov2018b,Fletcher2019,Jaskot2019,Plat2019}. 
The optical line equivalent widths of the most significant leakers (\fesc{}$>$20\%) are generally found to be very large, with [OIII]+H$\beta$ EWs $>$1500~\AA\ and H$\alpha$ EWs in excess of 900~\AA\ (e.g. \citealt{Izotov2016a,Izotov2016b,Izotov2018a,Izotov2018b,Vanzella2020}; c.f. \citealt{Shapley2016}), indicating that extremely large sSFR can be an effective phase in driving large escape fractions.
Sources with significant LyC leakage are also found to have compact sizes, exceptionally large O32 ratios ($>$6--30; \citealt{Izotov2018b}; \citealt{Jaskot2019}; \citealt{Nakajima2020}, \citealt{Vanzella2020}), and large EW \Lya{} emission (25--150 \AA{}) with a line profile indicating some mechanism for escape \citep[e.g.][]{Verhamme2015,Vanzella2016,RiveraThorsen2017,Marchi2018,Steidel2018,Fletcher2019}. 
While compact galaxies undergoing bursts of star formation may not prove to be the only mode of leaking significant fractions of ionizing radiation, it does appear that this is an effective pathway for galaxies to release LyC radiation to the IGM.  

The results presented in this paper demonstrate that a significant fraction of sources in the reionization era look remarkably like the LyC leakers found locally and at z$\simeq$3, with [OIII]+H$\beta$ equivalent widths seen in excess of 1200~\AA. 
Given trends at lower redshifts \citep{Tang2019,Du2020}, we expect that many of the  most extreme line emitters will also exhibit the O32 and \Lya{} properties seen in galaxies with large escape fractions.  
This is a significant departure from z$\simeq$2, where such properties are extremely rare in the galaxy population (e.g. Boyett et al. 2020, in prep).  

The increased detection rate of such extreme line emitting systems at z$\simeq$7 follows the evolution of the galaxy population toward more extreme sSFR activity. However, as is clear in Fig. \ref{fig:finalEWdistn}, the optical line EWs that are often linked to the largest escape fractions are well above the median of the [OIII]+H$\beta$ EW distribution at z$\simeq$7 (759~\AA; see \S3) and z$\simeq$8 (650--670~\AA; \citealt{Labbe2013,deBarros2019}).  
While clearly much work needs to be done in studying the range of galaxy properties that are linked to large LyC escape fractions, these results suggest that the conditions for large escape fractions may be optimized in the $\approx$20\% of z$\simeq$7 galaxies caught in the extreme emission line phase. 
When coupled with the large ionizing photon production efficiencies that accompanies such intense line emission (see above), it seems likely that this population of compact bursts with [OIII]+H$\beta$ EW$>$1200~\AA\ are among the most effective ionizing agents in the reionization era. 
While this is likely to be a short-lived phase, our results suggest that these objects become increasingly common at earlier times, potentially playing a very important role in reionization. 
Furthermore, our results indicate that conditions of high LyC escape and production likely exist in a significant subset of bright z$\simeq$7 galaxies, allowing for the possibility that luminous systems contribute significantly to reionization \citep{Naidu2020}. 
Future studies with \JWST{} will soon offer a much more detailed view of the demographics of the early galaxy population, enabling a more comprehensive comparison to LyC leaker samples that exist at lower redshifts. 

\section{Summary} \label{sec:summary}

We present a functional fit to the [OIII]$+$H$\beta$ EW distribution at z$\simeq$7, and use this fit to gain new insight into the ionizing output and cluster formation efficiency of galaxies during reionization. 
We infer the z$\simeq$7 [OIII]$+$H$\beta$ EW distribution from a sample of newly selected z$\simeq$6.63--6.83 galaxies where blue \Spitzer{}/IRAC [3.6]$-$[4.5] colours unambiguously indicate strong [OIII]$+$H$\beta$ emission.
To precisely identify galaxies in this redshift range, we develop a new colour selection which utilizes four filters at $\approx$1$\mu$m (including one narrow-band filter), yielding 20 bright (\Muv{} $\lesssim$ $-$21) galaxies over the wide-area COSMOS and XMM1 fields ($\sim$2.3 deg$^2$). 
To test whether the z$\simeq$7 [OIII]$+$H$\beta$ EW distribution evolves strongly with UV luminosity, we also infer the [OIII]$+$H$\beta$ EW distribution from a sample of 22 fainter (\Muv{} $\lesssim$ $-$20) \HST{}-selected galaxies at z$\sim$6.6--7.3 in the two GOODS fields.
In both sub-samples, we characterize the inferred z$\simeq$7 [OIII]$+$H$\beta$ EW distribution with a log-normal form, 
using two different methods to assess systematic uncertainties.  
Our conclusions are as follows:

\begin{enumerate}
    \item We infer the galaxy properties (e.g. stellar masses, ages, and [OIII]$+$H$\beta$ EWs) of all sources by fitting their photometry with a photoionization model using the BEAGLE SED fitting code. We infer [OIII]$+$H$\beta$ EWs for individual sources ranging from $\sim$200 to $\gtrsim$2000 \AA{} within both sub-samples. In general, sources with stronger IRAC colours, and hence larger inferred [OIII]$+$H$\beta$ EWs, have more precise inferred EWs with a median uncertainty of 0.17 dex amongst the strongest emitters (EW $>$1200 \AA{}). 
    
    \item As part of an ongoing spectroscopic campaign, we report a \Lya{} detection at z=6.850 (EW=15$\pm$3 \AA{}) from an extremely luminous (\Muv{} = $-$22.5$^{\scaleto{+0.1}{4.5pt}}_{\scaleto{-0.1}{4.5pt}}$) source with a very blue IRAC colour of [3.6]$-$[4.5] = $-$1.33$^{\scaleto{+0.26}{4.5pt}}_{\scaleto{-0.32}{4.5pt}}$. This IRAC colour implies extremely powerful [OIII]$+$H$\beta$ emission with EW = 4180$^{\scaleto{+1930}{4.5pt}}_{\scaleto{-1310}{4.5pt}}$ \AA{}. Such intense emission is indicative of extremely large sSFR (178$^{\scaleto{+238}{4.5pt}}_{\scaleto{-136}{4.5pt}}$ Gyr$^{-1}$) where the emergent starlight is dominated by a very young ($\sim$3 Myr) stellar population. Given its continuum 
    brightness (\textit{J}=24.45) and strong optical 
    nebular emission, this source is likely to be one of the best z$\simeq$7 targets for detailed rest-UV spectroscopic investigation. 
  
    \item We infer an [OIII]$+$H$\beta$ EW distribution with median EW = 759$^{\scaleto{+112}{4.5pt}}_{\scaleto{-113}{4.5pt}}$ \AA{} and standard deviation = 0.26$^{\scaleto{+0.06}{4.5pt}}_{\scaleto{-0.05}{4.5pt}}$ dex in our sample of luminous 
    z$\simeq$7 galaxies selected in COSMOS+XMM1.  We find no evidence that the z$\simeq$7 [OIII]$+$H$\beta$ EW distribution strongly evolves with UV luminosity when comparing to the GOODS sub-sample results. These estimates 
    of the average [OIII]+H$\beta$ are consistent within the uncertainties to those derived previously at z$\simeq$8 in the literature \citep{Labbe2013,deBarros2019}.  
    
    \item The strong [OIII]+H$\beta$ emission at z$\simeq$7 is a significant departure from z$\simeq$2, where typical massive galaxies are seen with [OIII]+H$\beta$ EW = 100--200~\AA.   This shift toward large nebular line EWs can be explained by evolution toward larger sSFR (4.4 Gyr$^{-1}$) and lower gas-phase metallicity (0.16 Z$_\odot$).  We also find evidence for the emergence of a population with 
    yet more intense nebular emission ([OIII]+H$\beta$ EW $>1200$~\AA) that is 
    very rarely seen in similar samples at lower redshifts.  
     These are compact 
    galaxies with very large sSFR ($>$30 Gyr$^{-1}$), implying rapid 
    mass doubling times and large star formation rate surface densities, as would be expected for galaxies caught in the midst of a burst of star formation.  While this is presumably a 
    short-lived phase, our results suggest that $\approx$20\% of the z$\simeq$7
    population possesses such intense nebular emission, implying 
    that galaxies likely cycle through this phase regularly in the reionization era.

    \item We argue that this population of extreme line emitters ([OIII]$+$H$\beta$ EW $>$ 1200~\AA) is likely to be very 
    effective at forming bound star clusters, based on trends 
    between cluster formation efficiency and star formation rate surface density found in nearby star forming galaxies (e.g. \citealt{Goddard2010,Adamo2011}). The rise in 
    abundance of this population of compact bursts at z$>$6 
    may signal a period of enhanced cluster formation in early galaxies.  
 
    \item We consider implications of our findings for the contribution of galaxies to reionization, leveraging recent insights from investigations of lower redshift populations.  
    We show that ionizing photon production is likely to be 
    enhanced relative to lower redshifts, with the most extreme 
    line emitting galaxies being the most prodigious in their 
    production efficiency.  We show that this population 
    of very extreme line emitters ([OIII]$+$H$\beta$ EW $>$1200~\AA) 
    is similar to the small samples of LyC leakers at lower redshift, 
    indicating that escape fractions may be maximized in this 
    subset of very large sSFR ($>$30 Gyr$^{-1}$) galaxies.  When coupled with their  
    large ionizing photon production efficiency, we suggest that this population 
    of compact bursts with [OIII]$+$H$\beta$ EW $>$1200~\AA\ may be among the most effective ionizing agents in the reionization era. 
  
\end{enumerate}

\section*{Acknowledgements}

We are grateful for insightful conversations with Eros Vanzella, and 
we thank Ramesh Mainali for assisting with our MMT/Binospec observations.  RE and DPS acknowledge funding from JWST/NIRCam contract to the University of Arizona, NAS5-02015. JC and SC acknowledge financial support from the European Research Council (ERC) via an Advanced Grant under grant agreement no. 321323 -- NEOGAL. Observations reported here were obtained at the MMT Observatory, a joint facility of the University of Arizona and the Smithsonian Institution.

This research has benefited from the SpeX Prism Library (and/or SpeX Prism Library Analysis Toolkit), maintained by Adam Burgasser at http://www.browndwarfs.org/spexprism. 
This research also made use of \textsc{astropy}, a community-developed core \textsc{python} package for Astronomy \citep{astropy:2013, astropy:2018}; \textsc{matplotlib} \citep{Hunter2007_matplotlib}; \textsc{numpy} \citep{van2011numpy}; and \textsc{scipy} \citep{jones_scipy_2001}.

\section*{Data Availability}
 
The optical through mid-infrared imaging data underlying this article are available through their respective data repositories. See \url{https://hsc-release.mtk.nao.ac.jp/doc/} for HSC data,  \url{http://www.eso.org/rm/publicAccess#/dataReleases} for UltraVISTA and VIDEO data, and \url{https://sha.ipac.caltech.edu/applications/Spitzer/SHA/} for IRAC data. Data will also be shared upon reasonable request to the corresponding author.




\bibliographystyle{mnras}
\bibliography{main} 



\appendix

\section{Rest-Optical Line Ratios} \label{appendix:line_ratios}

When calculating the \Spitzer{}/IRAC [3.6]$-$[4.5] colours as a function of [OIII]$+$H$\beta$ EW, we must determine the relative EW of each individual component ([OIII]$\lambda$4959, [OIII]$\lambda$5007, and H$\beta$) since lines will move in and out of the two IRAC bands at different redshifts. To do so, we adopt the theoretical [OIII]$\lambda$5007/[OIII]$\lambda$4959 line ratio of 2.98 \citep{StoreyZeippen2000}. We also adopt the empirical linear relation between H$\beta$ and [OIII]$\lambda$5007 EW obtained from the z$\simeq$2 results of \citet{Tang2019}:
\begin{equation}
    \logten{} \left(\mathrm{H}\beta\ \mathrm{EW}\right) = 1.065 \times \logten{} \left(\mathrm{[OIII]}\lambda5007\ \mathrm{EW}\right) - 0.938 
\end{equation}
All EWs are expressed in Angstroms.

We also seek to account for fainter emission lines when calculating the IRAC [3.6]$-$[4.5] colours. To do so, we consider the contribution from the Balmer lines H$\alpha$, H$\gamma$, and H$\delta$, calculating their EWs from H$\beta$ using the 10,000 K case B recombination ratios from \citet{OsterbrockFerland2006}. We also consider other moderately strong rest-optical metal lines that could fall into either IRAC band over the redshift range of interest. For [NII]$\lambda$6548,6583 and [SII]$\lambda$6716,6730, we use the theoretical 0.2 Z$_{\odot}$ line ratios from \citet{Anders2003}. For [OII]$\lambda$3727,3729, we fix the EW to 100 \AA{} for strong line emitters ([OIII]$\lambda$5007 EW $>$ 225 \AA{}) as motivated by the empirical results of \citet{Tang2019}. For weaker line emitters, we adopt the empirical linear relation with [OIII]$\lambda$5007 from \citet{Reddy2018_opticalLines}:
\begin{equation}
    \logten{} \left(\mathrm{[OII]}\ \mathrm{EW}\right) = 0.151 \times \logten{} \left(\mathrm{[OIII]}\lambda5007\ \mathrm{EW}\right) + 1.688
\end{equation}
Finally, we set the [NeIII]$\lambda$3869 EW equal to 0.5$\times$ the [OII] EW for extreme line emitters ([OIII]$\lambda$5007 EW $>$800 \AA{}) and 0.2$\times$ the [OII] EW for weaker emitters. This choice is motivated by the empirical results of \citet{Tang2019}.
 

\bsp	
\label{lastpage}
\end{document}